%
\catcode`@=11 
%
%

\font\fourteenrm=cmr10 scaled\magstep2
\font\twelverm=cmr10 scaled\magstep1
\font\ninerm=cmr9            \font\sixrm=cmr6

\font\fourteenbf=cmbx10 scaled\magstep2
\font\twelvebf=cmbx10 scaled\magstep1
\font\ninebf=cmbx9            \font\sixbf=cmbx6
\font\seventeeni=cmmi10 scaled\magstep3     \skewchar\seventeeni='177
\font\fourteeni=cmmi10 scaled\magstep2      \skewchar\fourteeni='177
\font\twelvei=cmmi10 scaled\magstep1        \skewchar\twelvei='177
\font\ninei=cmmi9                           \skewchar\ninei='177
\font\sixi=cmmi6                            \skewchar\sixi='177
\font\seventeensy=cmsy10 scaled\magstep3    \skewchar\seventeensy='60
\font\fourteensy=cmsy10 scaled\magstep2     \skewchar\fourteensy='60
\font\twelvesy=cmsy10 scaled\magstep1       \skewchar\twelvesy='60
\font\ninesy=cmsy9                          \skewchar\ninesy='60
\font\sixsy=cmsy6                           \skewchar\sixsy='60

\font\fourteenex=cmex10 scaled\magstep2
\font\twelveex=cmex10 scaled\magstep1

\font\fourteensl=cmsl10 scaled\magstep2                                         
\font\twelvesl=cmsl10 scaled\magstep1                                           
\font\ninesl=cmsl9                                                              

\font\fourteenit=cmti10 scaled\magstep2                                         
\font\twelveit=cmti10 scaled\magstep1                                           
\font\twelvett=cmtt10 scaled\magstep1                                           
\font\twelvecp=cmcsc10 scaled\magstep1                                          
\font\tencp=cmcsc10                                                             
\newfam\cpfam                                                                   
%
%
\newcount\f@ntkey            \f@ntkey=0                                         
\def\samef@nt{\relax \ifcase\f@ntkey \rm \or\oldstyle \or\or                    
         \or\it \or\sl \or\bf \or\tt \or\caps \fi }                             
\def\fourteenpoint{\relax                                                       
    \textfont0=\fourteenrm          \scriptfont0=\tenrm                         
    \scriptscriptfont0=\sevenrm                                                 
     \def\rm{\fam0 \fourteenrm \f@ntkey=0 }\relax                               
    \textfont1=\fourteeni           \scriptfont1=\teni                          
    \scriptscriptfont1=\seveni                                                  
     \def\oldstyle{\fam1 \fourteeni\f@ntkey=1 }\relax                           
    \textfont2=\fourteensy          \scriptfont2=\tensy                         
    \scriptscriptfont2=\sevensy                                                 
    \textfont3=\fourteenex     \scriptfont3=\fourteenex                         
    \scriptscriptfont3=\fourteenex                                              
    \def\it{\fam\itfam \fourteenit\f@ntkey=4 }\textfont\itfam=\fourteenit       
    \def\sl{\fam\slfam \fourteensl\f@ntkey=5 }\textfont\slfam=\fourteensl       
    \scriptfont\slfam=\tensl                                                    
    \def\bf{\fam\bffam \fourteenbf\f@ntkey=6 }\textfont\bffam=\fourteenbf       
    \scriptfont\bffam=\tenbf     \scriptscriptfont\bffam=\sevenbf               
    \def\tt{\fam\ttfam \twelvett \f@ntkey=7 }\textfont\ttfam=\twelvett          
    \h@big=11.9\p@{} \h@Big=16.1\p@{} \h@bigg=20.3\p@{} \h@Bigg=24.5\p@{}       
    \def\caps{\fam\cpfam \twelvecp \f@ntkey=8 }\textfont\cpfam=\twelvecp        
    \setbox\strutbox=\hbox{\vrule height 12pt depth 5pt width\z@}               
    \samef@nt}                                                                  
\def\twelvepoint{\relax                                                         
    \textfont0=\twelverm          \scriptfont0=\ninerm                          
    \scriptscriptfont0=\sixrm                                                   
     \def\rm{\fam0 \twelverm \f@ntkey=0 }\relax                                 
    \textfont1=\twelvei           \scriptfont1=\ninei                           
    \scriptscriptfont1=\sixi                                                    
     \def\oldstyle{\fam1 \twelvei\f@ntkey=1 }\relax                             
    \textfont2=\twelvesy          \scriptfont2=\ninesy                          
    \scriptscriptfont2=\sixsy                                                   
    \textfont3=\twelveex          \scriptfont3=\twelveex                        
    \scriptscriptfont3=\twelveex                                                
    \def\it{\fam\itfam \twelveit \f@ntkey=4 }\textfont\itfam=\twelveit          
    \def\sl{\fam\slfam \twelvesl \f@ntkey=5 }\textfont\slfam=\twelvesl          
    \scriptfont\slfam=\ninesl                                                   
    \def\bf{\fam\bffam \twelvebf \f@ntkey=6 }\textfont\bffam=\twelvebf          
    \scriptfont\bffam=\ninebf     \scriptscriptfont\bffam=\sixbf                
    \def\tt{\fam\ttfam \twelvett \f@ntkey=7 }\textfont\ttfam=\twelvett          
    \h@big=10.2\p@{}                                                            
    \h@Big=13.8\p@{}                                                            
    \h@bigg=17.4\p@{}                                                           
    \h@Bigg=21.0\p@{}                                                           
    \def\caps{\fam\cpfam \twelvecp \f@ntkey=8 }\textfont\cpfam=\twelvecp        
    \setbox\strutbox=\hbox{\vrule height 10pt depth 4pt width\z@}               
    \samef@nt}                                                                  
\def\tenpoint{\relax                                                            
    \textfont0=\tenrm          \scriptfont0=\sevenrm                            
    \scriptscriptfont0=\fiverm                                                  
    \def\rm{\fam0 \tenrm \f@ntkey=0 }\relax                                     
    \textfont1=\teni           \scriptfont1=\seveni                             
    \scriptscriptfont1=\fivei                                                   
    \def\oldstyle{\fam1 \teni \f@ntkey=1 }\relax                                
    \textfont2=\tensy          \scriptfont2=\sevensy                            
    \scriptscriptfont2=\fivesy                                                  
    \textfont3=\tenex          \scriptfont3=\tenex                              
    \scriptscriptfont3=\tenex                                                   
    \def\it{\fam\itfam \tenit \f@ntkey=4 }\textfont\itfam=\tenit                
    \def\sl{\fam\slfam \tensl \f@ntkey=5 }\textfont\slfam=\tensl                
    \def\bf{\fam\bffam \tenbf \f@ntkey=6 }\textfont\bffam=\tenbf                
    \scriptfont\bffam=\sevenbf     \scriptscriptfont\bffam=\fivebf              
    \def\tt{\fam\ttfam \tentt \f@ntkey=7 }\textfont\ttfam=\tentt                
    \def\caps{\fam\cpfam \tencp \f@ntkey=8 }\textfont\cpfam=\tencp              
    \setbox\strutbox=\hbox{\vrule height 8.5pt depth 3.5pt width\z@}            
    \samef@nt}                                                                  
%
%
%
%
\newdimen\h@big  \h@big=8.5\p@                                                  
\newdimen\h@Big  \h@Big=11.5\p@                                                 
\newdimen\h@bigg  \h@bigg=14.5\p@                                               
\newdimen\h@Bigg  \h@Bigg=17.5\p@                                               
\def\big#1{{\hbox{$\left#1\vbox to\h@big{}\right.\n@space$}}}                   
\def\Big#1{{\hbox{$\left#1\vbox to\h@Big{}\right.\n@space$}}}                   
\def\bigg#1{{\hbox{$\left#1\vbox to\h@bigg{}\right.\n@space$}}}                 
\def\Bigg#1{{\hbox{$\left#1\vbox to\h@Bigg{}\right.\n@space$}}}                 
%
%
%
\normalbaselineskip = 20pt plus 0.2pt minus 0.1pt                               
\normallineskip = 1.5pt plus 0.1pt minus 0.1pt                                  
\normallineskiplimit = 1.5pt                                                    
\newskip\normaldisplayskip                                                      
\normaldisplayskip = 20pt plus 5pt minus 10pt                                   
\newskip\normaldispshortskip                                                    
\normaldispshortskip = 6pt plus 5pt                                             
\newskip\normalparskip                                                          
\normalparskip = 6pt plus 2pt minus 1pt                                         
\newskip\skipregister                                                           
\skipregister = 5pt plus 2pt minus 1.5pt                                        
\newif\ifsingl@    \newif\ifdoubl@                                              
\newif\iftwelv@    \twelv@true                                                  
\def\singlespace{\singl@true\doubl@false\spaces@t}                              
\def\doublespace{\singl@false\doubl@true\spaces@t}                              
\def\normalspace{\singl@false\doubl@false\spaces@t}                             
\def\Tenpoint{\tenpoint\twelv@false\spaces@t}                                   
\def\Twelvepoint{\twelvepoint\twelv@true\spaces@t}                              
\def\spaces@t{\relax                                                            
      \iftwelv@ \ifsingl@\subspaces@t3:4;\else\subspaces@t1:1;\fi               
       \else \ifsingl@\subspaces@t3:5;\else\subspaces@t4:5;\fi \fi              
      \ifdoubl@ \multiply\baselineskip by 5                                     
         \divide\baselineskip by 4 \fi }                                        
\def\subspaces@t#1:#2;{                                                         
      \baselineskip = \normalbaselineskip                                       
      \multiply\baselineskip by #1 \divide\baselineskip by #2                   
      \lineskip = \normallineskip                                               
      \multiply\lineskip by #1 \divide\lineskip by #2                           
      \lineskiplimit = \normallineskiplimit                                     
      \multiply\lineskiplimit by #1 \divide\lineskiplimit by #2                 
      \parskip = \normalparskip                                                 
      \multiply\parskip by #1 \divide\parskip by #2                             
      \abovedisplayskip = \normaldisplayskip                                    
      \multiply\abovedisplayskip by #1 \divide\abovedisplayskip by #2           
      \belowdisplayskip = \abovedisplayskip                                     
      \abovedisplayshortskip = \normaldispshortskip                             
      \multiply\abovedisplayshortskip by #1                                     
        \divide\abovedisplayshortskip by #2                                     
      \belowdisplayshortskip = \abovedisplayshortskip                           
      \advance\belowdisplayshortskip by \belowdisplayskip                       
      \divide\belowdisplayshortskip by 2                                        
      \smallskipamount = \skipregister                                          
      \multiply\smallskipamount by #1 \divide\smallskipamount by #2             
      \medskipamount = \smallskipamount \multiply\medskipamount by 2            
      \bigskipamount = \smallskipamount \multiply\bigskipamount by 4 }          
\def\normalbaselines{ \baselineskip=\normalbaselineskip                         
   \lineskip=\normallineskip \lineskiplimit=\normallineskip                     
   \iftwelv@\else \multiply\baselineskip by 4 \divide\baselineskip by 5         
     \multiply\lineskiplimit by 4 \divide\lineskiplimit by 5                    
     \multiply\lineskip by 4 \divide\lineskip by 5 \fi }                        
\Twelvepoint  
\interlinepenalty=50                                                            
\interfootnotelinepenalty=5000                                                  
\predisplaypenalty=9000                                                         
\postdisplaypenalty=500                                                         
\hfuzz=1pt                                                                      
\vfuzz=0.2pt                                                                    
%
%
%
\def\pagecontents{                                                              
   \ifvoid\topins\else\unvbox\topins\vskip\skip\topins\fi                       
   \dimen@ = \dp255 \unvbox255                                                  
   \ifvoid\footins\else\vskip\skip\footins\footrule\unvbox\footins\fi           
   \ifr@ggedbottom \kern-\dimen@ \vfil \fi }                                    
\def\makeheadline{\vbox to 0pt{ \skip@=\topskip                                 
      \advance\skip@ by -12pt \advance\skip@ by -2\normalbaselineskip           
      \vskip\skip@ \line{\vbox to 12pt{}\the\headline} \vss                     
      }\nointerlineskip}                                                        
\def\makefootline{\baselineskip = 1.5\normalbaselineskip                        
                 \line{\the\footline}}                                          
\newif\iffrontpage                                                              
\newif\ifletterstyle                                                            
\newif\ifp@genum                                                                
\def\nopagenumbers{\p@genumfalse}                                               
\def\pagenumbers{\p@genumtrue}                                                  
\pagenumbers                                                                    
\newtoks\paperheadline                                                          
\newtoks\letterheadline                                                         
\newtoks\letterfrontheadline                                                    
\newtoks\lettermainheadline                                                     
\newtoks\paperfootline                                                          
\newtoks\letterfootline                                                         
\newtoks\date                                                                   
\footline={\ifletterstyle\the\letterfootline\else\the\paperfootline\fi}         
\paperfootline={\hss\iffrontpage\else\ifp@genum\tenrm\folio\hss\fi\fi}          
\letterfootline={\hfil}                                                         
\headline={\ifletterstyle\the\letterheadline\else\the\paperheadline\fi}         
\paperheadline={\hfil}                                                          
\letterheadline{\iffrontpage\the\letterfrontheadline                            
     \else\the\lettermainheadline\fi}                                           
\lettermainheadline={\rm\ifp@genum page \ \folio\fi\hfil\the\date}              
\def\monthname{\relax\ifcase\month 0/\or January\or February\or                 
   March\or April\or May\or June\or July\or August\or September\or              
   October\or November\or December\else\number\month/\fi}                       
\date={\monthname\ \number\day, \number\year}                                   
\countdef\pagenumber=1  \pagenumber=1                                           
\def\advancepageno{\global\advance\pageno by 1                                  
   \ifnum\pagenumber<0 \global\advance\pagenumber by -1                         
    \else\global\advance\pagenumber by 1 \fi \global\frontpagefalse }           
\def\folio{\ifnum\pagenumber<0 \romannumeral-\pagenumber                        
           \else \number\pagenumber \fi }                                       
\def\footrule{\dimen@=\prevdepth\nointerlineskip                                
   \vbox to 0pt{\vskip -0.25\baselineskip \hrule width 0.35\hsize \vss}         
   \prevdepth=\dimen@ }                                                         
\newtoks\foottokens                                                             
\foottokens={\Tenpoint\singlespace}                                             
\newdimen\footindent                                                            
\footindent=24pt                                                                
\def\vfootnote#1{\insert\footins\bgroup  \the\foottokens                        
   \interlinepenalty=\interfootnotelinepenalty \floatingpenalty=20000           
   \splittopskip=\ht\strutbox \boxmaxdepth=\dp\strutbox                         
   \leftskip=\footindent \rightskip=\z@skip                                     
   \parindent=0.5\footindent \parfillskip=0pt plus 1fil                         
   \spaceskip=\z@skip \xspaceskip=\z@skip                                       
   \Textindent{$ #1 $}\footstrut\futurelet\next\fo@t}                           
\def\Textindent#1{\noindent\llap{#1\enspace}\ignorespaces}                      
\def\footnote#1{\attach{#1}\vfootnote{#1}}

\let\footsymbol=\star                                                           
\newcount\lastf@@t           \lastf@@t=-1                                       
\newcount\footsymbolcount    \footsymbolcount=0                                 
\newif\ifPhysRev                                                                
\def\footsymbolgen{\relax \ifPhysRev \iffrontpage \NPsymbolgen\else             
      \PRsymbolgen\fi \else \NPsymbolgen\fi                                     
   \global\lastf@@t=\pageno \footsymbol }                                       
\def\NPsymbolgen{\ifnum\footsymbolcount<0 \global\footsymbolcount=0\fi          
   {\iffrontpage \else \advance\lastf@@t by 1 \fi                               
    \ifnum\lastf@@t<\pageno \global\footsymbolcount=0                           
     \else \global\advance\footsymbolcount by 1 \fi }                           
   \ifcase\footsymbolcount \fd@f\star\or \fd@f\dagger\or \fd@f\ast\or           
    \fd@f\ddagger\or \fd@f\natural\or \fd@f\diamond\or \fd@f\bullet\or          
    \fd@f\nabla\else \fd@f\dagger\global\footsymbolcount=0 \fi }                
\def\fd@f#1{\xdef\footsymbol{#1}}                                               
\def\PRsymbolgen{\ifnum\footsymbolcount>0 \global\footsymbolcount=0\fi          
      \global\advance\footsymbolcount by -1                                     
      \xdef\footsymbol{\sharp\number-\footsymbolcount} }                        
\def\space@ver#1{\let\@sf=\empty \ifmmode #1\else \ifhmode                      
   \edef\@sf{\spacefactor=\the\spacefactor}\unskip${}#1$\relax\fi\fi}           
\def\attach#1{\space@ver{\strut^{\mkern 2mu #1} }\@sf\ }                        
%
%
%
\newcount\chapternumber      \chapternumber=0                                   
\newcount\sectionnumber      \sectionnumber=0                                   
\newcount\equanumber         \equanumber=0                                      
\let\chapterlabel=0                                                             
\newtoks\chapterstyle        \chapterstyle={\Number}                            
\newskip\chapterskip         \chapterskip=\bigskipamount                        
\newskip\sectionskip         \sectionskip=\medskipamount                        
\newskip\headskip            \headskip=8pt plus 3pt minus 3pt                   
\newdimen\chapterminspace    \chapterminspace=15pc                              
\newdimen\sectionminspace    \sectionminspace=10pc                              
\newdimen\referenceminspace  \referenceminspace=25pc                            
\def\chapterreset{\global\advance\chapternumber by 1                            
   \ifnum\equanumber<0 \else\global\equanumber=0\fi                             
   \sectionnumber=0 \makel@bel}                                                 
\def\makel@bel{\xdef\chapterlabel{%
\the\chapterstyle{\the\chapternumber}.}}                                        
\def\sectionlabel{\number\sectionnumber \quad }                                 
\def\alphabetic#1{\count255='140 \advance\count255 by #1\char\count255}         
\def\Alphabetic#1{\count255='100 \advance\count255 by #1\char\count255}         
\def\Roman#1{\uppercase\expandafter{\romannumeral #1}}                          
\def\roman#1{\romannumeral #1}                                                  
\def\Number#1{\number #1}                                                       
\def\unnumberedchapters{\let\makel@bel=\relax \let\chapterlabel=\relax          
\let\sectionlabel=\relax \equanumber=-1 }                                       
\def\titlestyle#1{\par\begingroup \interlinepenalty=9999                        
     \leftskip=0.02\hsize plus 0.23\hsize minus 0.02\hsize                      
     \rightskip=\leftskip \parfillskip=0pt                                      
     \hyphenpenalty=9000 \exhyphenpenalty=9000                                  
     \tolerance=9999 \pretolerance=9000                                         
     \spaceskip=0.333em \xspaceskip=0.5em                                       
     \iftwelv@\fourteenpoint\else\twelvepoint\fi                                
   \noindent #1\par\endgroup }                                                  
\def\spacecheck#1{\dimen@=\pagegoal\advance\dimen@ by -\pagetotal               
   \ifdim\dimen@<#1 \ifdim\dimen@>0pt \vfil\break \fi\fi}                       
\def\chapter#1{\par \penalty-300 \vskip\chapterskip                             
   \spacecheck\chapterminspace                                                  
   \chapterreset \titlestyle{\chapterlabel \ #1}                                
   \nobreak\vskip\headskip \penalty 30000                                       
   \wlog{\string\chapter\ \chapterlabel} }                                      

\def\section#1{\par \ifnum\the\lastpenalty=30000\else                           
   \penalty-200\vskip\sectionskip \spacecheck\sectionminspace\fi                
   \wlog{\string\section\ \chapterlabel \the\sectionnumber}                     
   \global\advance\sectionnumber by 1  \noindent                                
   {\caps\enspace\chapterlabel \sectionlabel #1}\par                            
   \nobreak\vskip\headskip \penalty 30000 }                                     
\def\subsection#1{\par                                                          
   \ifnum\the\lastpenalty=30000\else \penalty-100\smallskip \fi                 
   \noindent\undertext{#1}\enspace \vadjust{\penalty5000}}                      

\def\undertext#1{\vtop{\hbox{#1}\kern 1pt \hrule}}                              
\def\APPENDIX#1#2{\par\penalty-300\vskip\chapterskip                            
   \spacecheck\chapterminspace \chapterreset \xdef\chapterlabel{#1}             
   \titlestyle{APPENDIX #2} \nobreak\vskip\headskip \penalty 30000              
   \wlog{\string\Appendix\ \chapterlabel} }                                     
\def\Appendix#1{\APPENDIX{#1}{#1}}                                              
\def\appendix{\APPENDIX{A}{}}                                                   
%
%
%
\def\eqname#1{\relax \ifnum\equanumber<0                                        
     \xdef#1{{\rm(\number-\equanumber)}}\global\advance\equanumber by -1        
    \else \global\advance\equanumber by 1                                       
      \xdef#1{{\rm(\chapterlabel \number\equanumber)}} \fi}                     
\def\eqinsert#1{\noalign{\dimen@=\prevdepth \nointerlineskip                    
   \setbox0=\hbox to\displaywidth{\hfil #1}                                     
   \vbox to 0pt{\vss\hbox{$\!\box0\!$}\kern-0.5\baselineskip}                   
   \prevdepth=\dimen@}}                                                         
%

%
                                                  
%
                                        
%
%
\def\GENITEM#1;#2{\par \hangafter=0 \hangindent=#1                              
    \Textindent{$ #2 $}\ignorespaces}                                           
\outer\def\newitem#1=#2;{\gdef#1{\GENITEM #2;}}                                 
\newdimen\itemsize                \itemsize=30pt                                
\newitem\item=1\itemsize;                                                       
\newitem\sitem=1.75\itemsize;                                
\newitem\ssitem=2.5\itemsize;                             
\outer\def\newlist#1=#2&#3&#4;{\toks0={#2}\toks1={#3}%
   \count255=\escapechar \escapechar=-1                                         
   \alloc@0\list\countdef\insc@unt\listcount     \listcount=0                   
   \edef#1{\par                                                                 
      \countdef\listcount=\the\allocationnumber                                 
      \advance\listcount by 1                                                   
      \hangafter=0 \hangindent=#4                                               
      \Textindent{\the\toks0{\listcount}\the\toks1}}                            
   \expandafter\expandafter\expandafter                                         
    \edef\c@t#1{begin}{\par                                                     
      \countdef\listcount=\the\allocationnumber \listcount=1                    
      \hangafter=0 \hangindent=#4                                               
      \Textindent{\the\toks0{\listcount}\the\toks1}}                            
   \expandafter\expandafter\expandafter                                         
    \edef\c@t#1{con}{\par \hangafter=0 \hangindent=#4 \noindent}                
   \escapechar=\count255}                                                       
\def\c@t#1#2{\csname\string#1#2\endcsname}                                      
\newlist\point=\Number&.&1.0\itemsize;                                          
\newlist\subpoint=(\alphabetic&)&1.75\itemsize;                                 
\newlist\subsubpoint=(\roman&)&2.5\itemsize;                                    
%

%
%
%
\newcount\referencecount     \referencecount=0                                  
\newif\ifreferenceopen       \newwrite\referencewrite                           
\newtoks\rw@toks                                                                
\def\NPrefmark#1{\attach{\scriptscriptstyle [ #1 ] }}
\let\PRrefmark\attach
\def\refmark#1{\relax\ifPhysRev\PRrefmark{#1}\else\NPrefmark{#1}\fi}            
\def\refend{\refmark{\number\referencecount}}                                   
\newcount\lastrefsbegincount \lastrefsbegincount=0                              
\def\refsend{\refmark{\count255=\referencecount                                 
   \advance\count255 by-\lastrefsbegincount                                     
   \ifcase\count255 \number\referencecount                                      
   \or \number\lastrefsbegincount,\number\referencecount                        
   \else \number\lastrefsbegincount-\number\referencecount \fi}}                
\def\ref#1{\REF\?{#1}\refend}

\def\refch@ck{\chardef\rw@write=\referencewrite                                 
   \ifreferenceopen \else \referenceopentrue                                    
   \immediate\openout\referencewrite=referenc.aux \fi}                     
\def\rw@begin#1\splitout{\rw@toks={#1}\relax                                    
   \immediate\write\rw@write{\the\rw@toks}\futurelet\n@xt\rw@next}              
\def\rw@next{\ifx\n@xt\rw@end \let\n@xt=\relax                                  
      \else \let\n@xt=\rw@begin \fi \n@xt}                                      
\let\rw@end=\relax                                                              
\let\splitout=\relax                                                            
\newdimen\refindent     \refindent=30pt                                         
\def\refitem#1{\par \hangafter=0 \hangindent=\refindent \Textindent{#1}}        
\def\REF#1#2{\space@ver{}\refch@ck                                              
   \global\advance\referencecount by 1 \xdef#1{\the\referencecount}%
   \immediate\write\referencewrite{\noexpand\refitem{#1.}}%
   \rw@begin #2\splitout\rw@end \@sf}                                           
\newcount\figurecount     \figurecount=0                                        
\newif\iffigureopen       \newwrite\figurewrite                                 
\def\figch@ck{\chardef\rw@write=\figurewrite \iffigureopen\else                 
   \immediate\openout\figurewrite=figures.aux                              
   \figureopentrue\fi}                                                          
\def\FIG#1#2{\figch@ck                                                          
   \global\advance\figurecount by 1 \xdef#1{\the\figurecount}%
   \immediate\write\figurewrite{\noexpand\refitem{#1.}}%
   \rw@begin #2\splitout\rw@end}                                                
%
                                                                            
%

%
%
%
\def\ugsfig#1 #2 #3 #4 {\midinsert
\centerline{
\special {insert #1}
\hbox spread #2 {}}
\vskip #3
\medskip
\centerline{Fig. #4}
\endinsert}
\newcount\tablecount     \tablecount=0                                          
\newif\iftableopen       \newwrite\tablewrite                                   
\def\tabch@ck{\chardef\rw@write=\tablewrite \iftableopen\else                   
   \immediate\openout\tablewrite=tables.aux                                
   \tableopentrue\fi}                                                           
\gdef\TABLE#1#2{\tabch@ck                                                       
   \global\advance\tablecount by 1 \xdef#1{\the\tablecount}%
   \immediate\write\tablewrite{\noexpand\refitem{#1.}}%
   \rw@begin #2\splitout\rw@end }                                               
                                            
%
                                                                            
%
%
%
\def\masterreset{\global\pagenumber=1 \global\chapternumber=0                   
   \global\equanumber=0 \global\sectionnumber=0                                 
   \global\referencecount=0 \global\figurecount=0 \global\tablecount=0 }        
\def\FRONTPAGE{\ifvoid255\else\vfill\penalty-2000\fi                            
      \masterreset\global\frontpagetrue                                         
      \global\lastf@@t=0 \global\footsymbolcount=0}                             

\def\paperstyle{\letterstylefalse\normalspace\papersize}                        
\def\letterstyle{\letterstyletrue\singlespace\lettersize}                       
\def\papersize{\hsize=35pc\vsize=50pc\hoffset=8pc\voffset=8pc                   
               \skip\footins=\bigskipamount}                                    
\def\lettersize{\hsize=6.5in\vsize=8.5in\hoffset=1in\voffset=1in
   \skip\footins=\smallskipamount \multiply\skip\footins by 3 }                 
\paperstyle   
%
%
\def\MEMO{\letterstyle\FRONTPAGE \letterfrontheadline={\hfil}                   
    \line{\quad\fourteenrm MEMORANDUM\hfil\twelverm\the\date\quad}         
    \medskip \memod@f}                                                          

\def\memit@m#1{\smallskip \hangafter=0 \hangindent=1in                          
      \Textindent{\caps #1}}                                                    
\def\memod@f{\xdef\to{\memit@m{To:}}\xdef\from{\memit@m{From:}}%
     \xdef\topic{\memit@m{Topic:}}\xdef\subject{\memit@m{Subject:}}%
     \xdef\rule{\bigskip\hrule height 1pt\bigskip}}                             
\memod@f                                                                        
%

%
%
%
%
\font\smallheadfont=cmss10 at 10truept
\font\largeheadfont=cmr10 at 20.74truept 
\newdimen\longinden \longinden=0.5 truein	
\newdimen\leftmarg \leftmarg=\hoffset \advance \leftmarg by -.5in
\newdimen\tmp
\newbox\headbox
\newbox\tmpbox
\setbox\headbox=
   \vbox {\offinterlineskip	
      \hbox to7.5in{}
      \vbox {
        \hbox {\hskip\longinden {\largeheadfont Northeastern University}}}
      \vskip 3truept
      \hrule height2pt \vskip1pt \hrule 
      \vskip 5truept
      \vbox {
        \hbox {\hskip\longinden {\smallheadfont 360 Huntington Avenue, Boston,
      Massachusetts, 02115}}}
      \vskip 9truept
      \hrule
      \vskip 5truept
      \vbox {
        \hbox {\hskip\longinden {\smallheadfont Department of Physics}}
        \vskip 3.5pt \hbox {\hskip\longinden 
{\smallheadfont J. B. Sokoloff
\hskip3em\relax  Tel. (617) 373-2931  \hskip3em\relax 
Internet: 3630jbs@neu.edu \hskip3em\relax FAX (617) 373-2943}}}
      \vskip 20truept
      \hrule \vskip1pt \hrule height2pt
    }
\tmp=\ht\headbox
\advance \tmp by -\voffset	
\advance \tmp by 0.5truein
\setbox\tmpbox=\vbox to \tmp{\vss \box\headbox}
\def\depthead{\vskip-\lastskip
    \moveleft\leftmarg\box\tmpbox\bigskip}
\newskip\lettertopfil                                                           
\lettertopfil = 0pt plus 1.5in minus 0pt                                        
\newskip\letterbottomfil                                                        
\letterbottomfil = 0pt plus 2.3in minus 0pt                                     
\newskip\spskip \setbox0\hbox{\ } \spskip=-1\wd0                                
\def\addressee#1{\medskip\rightline{\the\date\hskip 30pt} \bigskip              
   \vskip\lettertopfil                                                          
   \ialign to\hsize{\strut ##\hfil\tabskip 0pt plus \hsize \cr #1\crcr}         
   \medskip\noindent\hskip\spskip}                                              
\newskip\signatureskip       \signatureskip=40pt                                
\def\signed#1{\par \penalty 9000 \bigskip \dt@pfalse                            
  \everycr={\noalign{\ifdt@p\vskip\signatureskip\global\dt@pfalse\fi}}          
  \setbox0=\vbox{\singlespace \halign{\tabskip 0pt \strut ##\hfil\cr            
   \noalign{\global\dt@ptrue}#1\crcr}}                                          
  \line{\hskip 0.5\hsize minus 0.5\hsize \box0\hfil} \medskip }                 

\def\endletter{\ifnum\pagenumber=1 \vskip\letterbottomfil\supereject            
\else \vfil\supereject \fi}                                                     
\newbox\letterb@x                                                               
\def\lettertext{\par\unvcopy\letterb@x\par}                                     
\def\multiletter{\setbox\letterb@x=\vbox\bgroup                                 
      \everypar{\vrule height 1\baselineskip depth 0pt width 0pt }              
      \singlespace \topskip=\baselineskip }                                     
\def\letterend{\par\egroup}                                                     
%
%
%
\newskip\frontpageskip                                                          
\newtoks\pubtype                                                                
\newtoks\Pubnum                                                                 
\newtoks\pubnum                                                                 
\newif\ifp@bblock  \p@bblocktrue                                                
\def\PH@SR@V{\doubl@true \baselineskip=24.1pt plus 0.2pt minus 0.1pt            
             \parskip= 3pt plus 2pt minus 1pt }                                 
\def\PHYSREV{\paperstyle\PhysRevtrue\PH@SR@V}                                   
\def\titlepage{\FRONTPAGE\paperstyle\ifPhysRev\PH@SR@V\fi                       
   \ifp@bblock\p@bblock\fi}                                                     
\def\nopubblock{\p@bblockfalse}                                                 
\def\endpage{\vfil\break}                                                       
\frontpageskip=1\medskipamount plus .5fil                                       
\pubtype={\tensl Preliminary Version}                                           
\Pubnum={$\caps NUB - \the\pubnum $}                                     
\pubnum={0000}                                                                  
\def\p@bblock{\begingroup \tabskip=\hsize minus \hsize                          
   \baselineskip=1.5\ht\strutbox \topspace-2\baselineskip                       
   \halign to\hsize{\strut ##\hfil\tabskip=0pt\crcr                             
   \the\Pubnum\cr \the\date\cr \the\pubtype\cr}\endgroup}                       
\def\title#1{\vskip\frontpageskip \titlestyle{#1} \vskip\headskip }             
\def\author#1{\vskip\frontpageskip\titlestyle{\twelvecp #1}\nobreak}

\def\address#1{\par\kern 5pt\titlestyle{\twelvepoint\it #1}}                    
\def\andaddress{\par\kern 5pt \centerline{\sl and} \address}                    

\def\abstract{\vskip\frontpageskip\centerline{\fourteenrm ABSTRACT}             
              \vskip\headskip }

%
%
%

\def\\{\relax\ifmmode\backslash\else$\backslash$\fi}                            
\def\globaleqnumbers{\relax\if\equanumber<0\else\global\equanumber=-1\fi}       
\def\nextline{\unskip\nobreak\hskip\parfillskip\break}                          
                       
\def\journal#1&#2(#3){\unskip, \sl #1~\bf #2 \rm (19#3) }                       

\def\topspace{\hrule height 0pt depth 0pt \vskip}                               

\let\int=\intop                                                
\def\prop{\mathrel{{\mathchoice{\pr@p\scriptstyle}{\pr@p\scriptstyle}{          
                \pr@p\scriptscriptstyle}{\pr@p\scriptscriptstyle} }}}           
\def\pr@p#1{\setbox0=\hbox{$\cal #1 \char'103$}                                 
   \hbox{$\cal #1 \char'117$\kern-.4\wd0\box0}}                                 
\def\lsim{\mathrel{\mathpalette\@versim<}}                                      
\def\gsim{\mathrel{\mathpalette\@versim>}}                                      
\def\@versim#1#2{\lower0.2ex\vbox{\baselineskip\z@skip\lineskip\z@skip          
  \lineskiplimit\z@\ialign{$\m@th#1\hfil##\hfil$\crcr#2\crcr\sim\crcr}}}        
%
%
%
\let\sec@nt=\sec                                                                
\def\sec{\relax\ifmmode\let\n@xt=\sec@nt\else\let\n@xt\section\fi\n@xt}         
\def\obsolete#1{\message{Macro \string #1 is obsolete.}}                        
\def\firstsec#1{\obsolete\firstsec \section{#1}}                                
\def\firstsubsec#1{\obsolete\firstsubsec \subsection{#1}}                       
\def\thispage#1{\obsolete\thispage \global\pagenumber=#1\frontpagefalse}        
\def\thischapter#1{\obsolete\thischapter \global\chapternumber=#1}              
\def\nextequation#1{\obsolete\nextequation \global\equanumber=#1                
   \ifnum\the\equanumber>0 \global\advance\equanumber by 1 \fi}                 
\def\BOXITEM{\afterassigment\B@XITEM\setbox0=}                                  
\def\B@XITEM{\par\hangindent\wd0 \noindent\box0 }                               
%
%
\catcode`@=12 
%
%

\hoffset=0.5in
\voffset=0.5in
\overfullrule=0pt
\baselineskip=15pt
\centerline{\bf Explaining the Virtual Universal Occurrence 
of Static Friction} 
\centerline{J. B. Sokoloff$^*$, Physics Department and}
\centerline{Center for Interdisciplinary Research on complex Systems,}
\centerline{Northeastern University, Boston, Massachusetts 02115}
\medskip
\centerline{abstract}

Perturbation theory, simulations and scaling arguments predict 
that there should be no static friction for 
two weakly interacting flat atomically smooth clean solid surfaces. 
The absence of static friction results from the fact that the atomic level 
interfacial potential energy is much weaker than the elastic potential energy, 
which prevents the atoms from sinking to their interfacial 
potential minima. Consequently, we have essentially two rigid solids, for 
which the forces at randomly distributed "pinning sites" cancel. It is 
shown here that even fluctuations in the 
concentration of atomic level defects at the interface do not account for 
static friction. 
It is also argued that the sliding of contacting asperities, which occurs 
when the problem is studied at the multi-micron length scale, relative to 
each other involves the shearing of planes of atoms. 
Since this results in a force for the interaction of two asperities which 
varies over sliding distances of the order of an atomic spacing, the 
contacting asperities at the surface are able to 
sink to their interfacial potential minima, with negligible cost in elastic 
potential energy. This results in static friction. 
\smallskip
\noindent {\bf Keywords:} Semi-empierical and model calculations, friction, 
surface defects.
\bigskip
\noindent * Corresponding author-address: Physics Department, Northeastern 
University, Boston, Massachusetts 02115. Phone number: (617) 373-2931, FAX 
number: (617) 373-2943, E-mail address: 3630jbs@neu.edu. 
\endpage
\noindent {\bf I. Introduction}

It is well known to every student in an elementary physics class that kinetic 
friction has very little velocity dependence in the slow sliding speed limit 
(often called "dry friction'). Molecular dynamics simulations[1] and 
analytic calculations[2,3] show that while commensurate interfaces are 
pinned for applied forces below a critical value (i.e., exhibit static 
friction), incommensurate surfaces are not pinned and exhibit viscous friction 
(i.e., friction proportional to the sliding velocity) for sufficiently weak 
interfacial forces.
Perturbation theory calculations done for a nonmetallic monolayer film with 
underdamped 
phonons sliding on a nonmetallic substrate with some disorder, however, give 
nearly velocity independent sliding friction[4] and exhibit a divergence 
in the mean square atomic displacement in the limit of zero sliding velocity. 
The latter behavior signifies that the film will be pinned below a critical 
applied force. This behavior has been confirmed by recent molecular dynamics 
calculations on such a system[5]. Perturbation theory calculations done 
for a three dimensional film sliding on a substrate, however, give viscous 
friction. This result is consistent with the notion 
that without multistability, there cannot be "dry friction" due 
to vibrational excitations in an elastic solid[6,7]. Dry friction is 
possible for the monolayer film, as mentioned above, however, because the two 
dimensional phonon density of states of the film does not drop to zero as 
the frequency goes to zero, as it does for a three dimensional solid. 
As the sliding velocity drops to zero, the "washboard frequency" (the sliding 
velocity divided by a lattice constant) drops to zero. Since the phonon density 
of states does not drop to zero, there are phonons present at arbitrarily low 
frequency, which can be excited by the substrate potential. Since the density 
of states does fall to zero as the frequency falls to zero in three dimensions, 
however, the force of friction falls to zero as the velocity does. 
In models used for charge density waves (CDW), in which the CDW is modeled as 
an elastic medium moving through a solid containing impurities distributed 
randomly throughout it, there is no pinning in 
four or more dimensions[8]. In contrast, in fewer than four dimensions, there 
is pinning. For models used for friction[9], consisting of a three 
dimensional elastic medium moving over a surface containing a random array 
of point defects, the critical dimension is three[9]. As a consequence, 
although
if the defect potential is sufficiently large, there will be static friction 
and "dry friction," for weak defect potentials, there will be no static 
friction and the kinetic friction will be viscous (i.e., linear in 
the sliding velocity). The non-periodic "defect potential" acting across the 
interface could be due to disorder on any length scale in the problem. For 
example, it can be due to atomic level point defects, such as vacancies and 
substitutional 
impurities at the interface, as has been assumed in references 4 and 5, but it can 
also be due to the fact that the surfaces of the sliding solids are only 
in contact at micron scale randomly located protrusions, commonly known as 
asperities. On the atomic scale, it can also be due to adsorbed film 
molecules[2]. 

In contrast to atomic level point defects, however, asperities and 
adsorbed molecules possess internal structure and as a consequence if they are 
sufficiently flexible, they can exhibit multistability (i.e., the existence 
of more than one stable configuration needed for Tomlinson model[6] to apply). 
Caroli and Noziere[7] proposed an 
explanation for "dry friction" based on the Tomlinson model[6]. 
In the Tomlinson model the 
two bodies which are sliding relative to each other at relatively slow speeds 
remain stuck together locally until their centers of mass have slid a small 
distance relative to
each other, at which point the stuck configuration of the two surfaces becomes 
unstable and the two surfaces locally slip rapidly with respect to each other 
until they become stuck again, and the process repeats itself. The slipping 
motion 
that takes place can either be local or can involve motion of the body as a 
whole. Then the actual friction acting locally at the interface could 
be viscous, but the rapid motion that takes place, even at slow 
sliding speeds, could still result in 
a sizable amount of friction, even in the limit of vanishing average sliding 
velocity. In Caroli and 
Noziere's model[7] interface contact only occurred at a very 
dilute concentration of interlocking asperities. It is the rapid stick-slip 
motion of these asperities, which gives rise to dry friction on the average 
in this model, 
if we assume that all of the kinetic energy released in the slip is dissipated, 
which is probably true in the zero velocity limit. 
This mechanism would seem to imply that the occurrence of dry friction depends 
on the existence of multistability; 
in situations in which the asperities are not multistable, there will be 
neither "dry friction" nor static friction. It is argued[7] that in the absence 
of plasticity 
in the model, the maximum force of dynamic friction, in the velocity approaches 
zero limit, must be equal to the force of static friction. 
It was pointed out in recent work by these 
workers that the asperities are generally too stiff to undergo the Tomlinson 
model-like instabilities because of their shape[7]. Therefore, these workers 
proposed alternative mechanisms. In one mechanism, 
it is assumed that there exists a glassy film at the interface in 
the experimental systems that they are trying to describe. Glassy materials 
possess metastable atomic configurations (the equivalent of "two-level systems" 
which are believed to contribute to the specific heat) which can exhibit 
Tomlinson model-like instabilities during sliding, similar to those found by 
Falk and Langer in their study of the shearing of glassy materials[10]. 
This mechanism will, however, only be the correct explanation of "dry friction" for 
glassy interfaces. It is not clear, however, that all interfaces 
are glassy. He, et. al.[2], proposed that the occurrence of static 
friction between elastic solids requires the existence of an adsorbed 
sub-monolayer film 
at the interface. It is important to know if the occurrence of a glassy 
interface or adsorbed molecules is a requirement for the occurrence of static 
friction. If it is, it would imply that clean interfaces between crystalline 
solids would not exhibit static friction. Caroli and Noziere[7] 
also proposed that adhesive forces could provide the 
required multistability (because of the so called "jump to contact" instability) 
to give friction at slow speed. It is not clear that this will be significant 
for asperities under load, however. In the absence of multistability, there 
is good reason to believe that there will be neither static friction nor dry 
friction, at least for light enough loads to put us in the "weak pinning limit" 
in the language of the charge density wave and vortex problems first studied 
by Larkin and Ovchinikov[11] and Fukuyama, Lee and Rice[12]. The existence 
of multistability has also been shown to be a condition for pinning of CDW's 
$^1$ [13]. [One way of
understanding this is that if there is static friction, the sliding velocity
of the solid will only be nonzero if a force above the force of static
friction is applied to the body. Alternatively, we can view this as implying
that the force of friction approaches a nonzero value as the center of mass
velocity approaches zero. The arguments in this reference and Ref. 7 tell us
that there must be multistability for this to occur.]

In section II, a discussion is 
given of the scaling theoretic treatment of static friction. This is an 
outgrowth of a similar treatment by Fisher of the pinning of charge density 
waves. It is found that, at least for surfaces with defects that produce a 
relatively weak potential, the Larkin domains (i.e., the regions over which 
the solid distorts to accommodate defects at the interface) are as large 
as the interface, as was found by Persson and Tosatti using perturbation[9] 
theory. This implies that the static friction decreases as the inverse of the 
square root of the interface area, as was found for perfect crystalline 
interfaces by Muser, et. al.[2]. In section, III, it is shown that when one takes 
into account the distribution of contacting asperities at the interface, 
one finds that the asperities are in the "strong pinning limit," in which the 
Larkin domains are much smaller than the interface, implying that 
there is static friction. This is shown to be a consequence of the fact that 
the shear force between two contacting asperities varies as the asperities 
are slipped relative to each other over slip distances of the order of 
atomic spacings, resulting in a force constant much larger than that due 
to the elastic force constant of the asperities. As a consequence, 
the asperities satisfy the criterion for the occurrence of multistability, 
shown in Ref. 7 to be a requirement for the occurrence of static 
friction. 
\endpage

\noindent {\bf II. Scaling Treatment at the Atomic Level of Static 
Friction}

In this section, we will treat the problem of static 
friction due to disorder which results from atomic level defects, such as 
vacancies or substitutional impurities using scaling arguments. 
In the next section, we will consider random contacting 
asperities, which occur when the surface is viewed on the micron length scale.

Following Fisher's treatment of the charge density wave (CDW) problem[8], it is 
clear that we can also use a scaling argument for the friction problem in 
order to determine whether the pinning potential becomes irrelevant as the 
length scale becomes large. In order to accomplish this, let us formulate 
this problem in a way similar to the 
way that Fisher does, by considering the crystal lattice to be sliding over 
a disordered substrate potential under the influence of a force F, which is 
applied to each atom in the crystal. Then we can write the equation of motion 
as 
$$m{\bf \ddot u}+m\gamma {\bf\dot{u}}_j=
\sum_{j'} {\bf D} ({\bf R}_j-{\bf R}_{j'})\cdot {\bf u}_{j'}
-{\bf f}({\bf R}_j)+{\bf F}, \eqno (1)$$
where ${\bf D} ({\bf R}_j-{\bf R}_{j'})$ is the force constant matrix for 
the lattice, ${\bf f}({\bf R_j})$ is the force due to the substrate on the 
$j^{th}$ atom and ${\bf R}_j$ is the location of the $j^{th}$ atom in the 
lattice. As a result of slow speed sliding of the 
lattice over the disordered substrate potential, low frequency acoustic 
phonons are excited. Since these modes have wavevector ${\bf k}$, small 
compared to the Brillouin zone radius, ${\bf u}_j$, 
the displacement of the $j^{th}$ atom is a slowly varying function of 
${\bf R}_j$. Then, following the discussion in Ref. 14, we can write the 
first term on the right hand side of Eq. (1) as 
$${\bf D'} (i^{-1}\nabla_j) {\bf u}_j, \eqno (2)$$
where ${\bf D'} ({\bf k})$ is the Fourier transform of 
${\bf D} ({\bf R}_j-{\bf R}_{j'})$ and 
$\nabla_j=(\partial/\partial X_j,\partial/\partial Y_j,\partial/\partial Z_j)$ 
where ${\bf R}_j=(X_j,Y_j,Z_j)$ to a good approximation. Furthermore, 
to a good approximation we can expand ${\bf D'}$ to second order in 
$\nabla_j$. Equation (1) then becomes
$$m{\bf \ddot u}+m\gamma {\bf\dot{u}}_j=
-vE' \nabla_j^2{\bf u}_{j'}
-{\bf f}({\bf R}_j)+{\bf F}, \eqno (3)$$
where E' is an effective Young's modulus and v is the unit cell volume. 
We can then apply Fisher's scaling argument[8] to the 
resulting equation. This is accomplished by dividing the solid into blocks of 
of length L lattice sites parallel to the interface and L' lattice sites 
normal to the 
interface, assuming that these dimensions are chosen so that ${\bf u}_{j}$ 
varies slowly over each such a block. Then integrating equation (15) over a 
block which lies at the interface, we obtain
$$L^2 L'[m{\bf \ddot u}_{j'}+m\gamma {\bf\dot{u}}_{j'}]=
-(1/2)L'L^2(E'/L^2) {\nabla'^2}_{j} {\bf u}_{j'}
-L{\bf f'}({\bf R'}_j)+L^2 L'{\bf F},\eqno (4)$$
where ${\nabla'}_j^2$ denotes the Laplacian in the rescaled 
coordinates $[(X'_{j'},Y'_{j'},Z'_{j'})=(X_j/L,Y_j/L,Z_j/L')]$, which are 
the coordinates of the centers of the boxes. Here we made use of the fact that 
${\bf u}_j$ varies on length scales L and L', when we transform to the 
block coordinates. Hence, we may replace $\nabla^2_j$ by $L^{-2}\nabla'^2_{j}$ 
and ${\bf f}({\bf R}_j)$ by $L{\bf f'}({\bf R}_{j'})$
The substrate force is only 
multiplied by L because the interaction of the defects with a single block at 
the surface is proportional to the square root of the surface area of the 
block.  For a thick solid (i.e., one whose 
lateral and transverse dimensions are comparable), L and L' are always of 
comparable magnitude. Then, we conclude that no matter how large we make 
L and L',
 the ratio of the elastic force (the first term on the right hand side) and the
substrate force (the second term on the right hand side) will remain the same. 
This implies that we are at the critical dimension for this problem since when 
the length scales L and L' are increased, neither the elasticity nor the 
substrate force becomes irrelevant. Which ever one dominates at one length scale 
will dominate at all. 
Eq. (4) implies that the force of static friction per unit area acting at the 
interface is inversely proportional to the square root of the interface area.

An alternative way to consider static friction due to Larkin domains is to 
minimize the potential energy of the solid in contact with the substrate with 
respect to the size of a Larkin domain[11,12]. Given that the energy density 
of the elastic solid is given by
$$(1/2)E' |\nabla {\bf u}|^2+V({\bf R}_j)\delta (z), \eqno (5)$$
where $V({\bf R}_j)$ is the potential per unit area,  
the energy of a single Larkin 
domain is given by
$$E=(1/2)L'L^2 vE' [|\nabla'_t u|^2/L^2+|\partial u/\partial z'|^2/L'^2]-V_0
L, \eqno (6)$$
where v is the crystal unit cell volume and $V_0$ is a typical value of 
the potential energy, 
which when minimized gives that $L'\approx L.$ Then the energy per unit area 
within a Larkin domain height of the interface is given by
$$E/L^2\approx [(1/2)vE' |\nabla' u|^2-V_0]/L \eqno(7)$$
whose absolute minimum occurs for infinite L (more correctly L comparable to 
the interface length in units of a lattice constant) for 
$E' |\nabla' u|^2>V_0$, implying that the 
Larkin domain energy is minimized when L is equal to the size of the elastic 
solid in units of a lattice constant. Then, the sliding 
elastic solid behaves as a rigid solid, and hence the static friction per 
unit area decreases as the reciprocal of the square root of the surface area. 

If the sliding solid is thin (i.e. has dimensions normal to the interface much 
smaller than those along it), the L' can only be made as large as the thickness 
in units of a lattice constant, 
but L can have any value. In this case, we conclude from equation 
(6) 
that the substrate force dominates over the elastic forces once L becomes 
comparable to the thickness. This means that there are now Larkin domains of 
size comparable to the ratio of elastic to substrate force times the thickness. 
The interface will then consist of many Larkin domains. Since the pinning force 
(i.e., static friction) scales with the number of Larkin domains on the 
interface, this implies that the force of static friction per unit area approaches 
a non-zero value. This argument is consistent with our conclusion in the last 
section, based on perturbation theory, that a thin solid exhibits static 
friction, but a thick solid will not. 

Next, let us consider the effect of fluctuations in the defect concentration, 
for a thick solid. To do this, we again divide the solid into boxes of length L 
and examine what percentage of the blocks at the interface contain a large 
enough concentration of defects to put these blocks in the 
"strong pinning" regime (i.e., the regime in which the substrate forces 
dominate over the elastic forces between the blocks). To examine whether 
this is possible, let us define a parameter $\lambda=V_1/E'b^3$, 
where b is a lattice 
spacing and $V_1$ is the strength of the potential due to a defect acting 
on an atom. Let $n_c=c'L^2$ be the number of defects within a particular 
strongly pinned block and c' is the defect concentration large enough 
for it to be considered a strongly pinned block, i. e. a block whose 
interaction with the substrate is much 
larger than the elastic interaction between such blocks. (This concentration 
is necessarily noticeably larger than the mean defect 
concentration on the interface.) Then the interaction of a typical strong 
block with the substrate defects is $\lambda (c'L^2)^{1/2}$. The interface 
area surrounding each strong block is the total interface area A divided by 
the number of strong blocks at the interface, which is equal to $PA/(b L)^2$, 
where P is the probability of a particular block being a strong one [$A/(bL)^2$ 
is clearly equal to the total number of blocks at the interface, both 
strong and weak]. Then we obtain $(bL)^2/P$ for the area surrounding a strongly 
pinned block. Then the typical length for the 
elastic energy acting between two strong blocks is L' b where $L'=L/P^{1/2}$. 
The ratio of the total elastic energy associated with each strong 
block to $E' b^3$ is the product of the volume surrounding a strongly pinned 
block=$(L')^3$ with $(L')^{-2}$, since $\nabla^2 u\propto L'^{-2}$, or 
L'. Then the criterion for a block to be a strong block is 
$\lambda (c'L^2)^{1/2}>>L'$, or $\lambda>>(c' P)^{-1/2}$. Since $c' P<1$, this violates 
our previous assumption that $\lambda<<1$. Thus, we conclude that such 
fluctuations in the defect concentration will not lead to static friction. 

If we assume that the sliding solid and the substrate surfaces at the interface 
are incommensurate and that the defects are either vacancies or substitutional 
impurities, which are centered around particular lattice sites in the 
substrate, 
there is another type of concentration fluctuation. For a uniform random 
distribution of defects over the substrate lattice sites, surface atoms 
of a completely rigid sliding solid will be found at all possible position 
within the various defect 
potential wells, which results in the net force on the solid due to defects 
being zero on the average. Let us again divide the solid into blocks of 
length L, but now the concentration of defects in each block will be taken 
to be equal to the mean defect concentration c. We will look for blocks in 
which the defects are distributed such that there is a sizable concentration 
of atoms located in that region of defect potentials, for which the 
force on the block is opposite the direction in which we are 
attempting to slide the block, due to the defects. Then we can divide each 
block into regions of equal size. If a defect lies in one region, the atoms 
which interact with the defect will have a force exerted on them opposing the 
attempt to slide the solid, and if it lies in the other region, the force 
on the atoms that it interacts with will be in the opposite direction. These 
two regions are, of course, fragmented. Then the net force on a block of 
length L at the interface due to the substrate is proportional to $L^2 \delta 
c P$, where $\delta c$ is the mean difference in defect concentrations between 
the two regions in the block defined above (such that the mean defect 
concentration over the whole block is c) and P is the probability of having 
a concentration difference $\delta c$ between these two regions. The 
un-normalized probability of having a concentration difference $\delta c$ 
between the two regions in a block is given by
$${N_1 !\over n_{c1}! ({1\over 2}N_1-n_{c1})! (cN_1-n_{c1})!
({1\over 2}N_1-cN_1+n_{c1})!}, \eqno (8)$$
where the number of atoms in a block $N_1=L^2$ and and the number of atoms in 
the region in which the net force is against the direction in which we are 
attempting to slide the solid $n_{c1}=(1/2)(c+(1/2)\delta c)N_1$, 
whose normalized small $\delta c$ approximation is
$$P\approx exp(-{3N_1\over 2c(1-c)}\delta c^2). \eqno (9)$$
Since in the large $N_1$ limit P decreases exponentially with increasing $N_1,$ 
unless 
$\delta c\approx N_1^{-1/2}$, we conclude that the substrate defect force on 
the block will not increase with increasing $N_1$ and hence in the large 
$N_1$ limit, the elastic force on the block, which we showed above is 
proportional to L=$N_1^{1/2}$, will dominate. This implies that this type 
of concentration fluctuation will not lead to static friction for a 
macroscopic size block. 

Although fluctuations in the defect concentration will not lead to static 
friction for defects whose potential strength is small compared to typical 
interatomic elastic energies, fluctuations in the defect strength can, if the 
defect strength distribution contains fluctuations which are larger than 
interdefect elastic forces. Let us now consider this possibility. 
The energy of the interface consists of two parts. One part is the single
defect energy, which consists of the interaction energy of a defect 
with the substrate plus the elastic energy cost necessary for
each defect to seek its minimum energy, neglecting its elastic interaction
with other defects, which is independent of the defect density. 
It should be noted that there is a restoring force when the defect 
is displaced relative to the center of mass, even if the defect-defect 
interaction is neglected[7,15].
The second part is the elastic interaction between 
defects within the same solid, which depends on the defect density.
In order to determine the effect of these energies,
let us for simplicity model the interaction of the $\ell^{th}$ defect 
with the lattice by a
spherically symmetric harmonic potential of force constant $\alpha_{\ell}$.
Assume that
in the absence of distortion of the solid, the $\ell^{th}$ defect 
lies a distance
${\bf \Delta}_{\ell}$ from the minimum of its potential
well. Let ${\bf u}_{\ell}$ be the displacement of the $\ell^{th}$
defect from its initial position.
We use the usual elastic Green's function tensor of the medium at a distance
r from the point at which a force is applied at the interface,
but for simplicity, we approximate it by
the simplified form $G(r)=(E'r)^{-1}$, where E' is
Young's modulus[15]. Then the
equilibrium conditions on the u's are
$${\bf u}_{\ell}=(E'a)^{-1}\alpha_{\ell} ({\bf \Delta}_{\ell}-{\bf u}_{\ell})+
\sum_j (E'R_{\ell,j})^{-1}
\alpha_j ({\bf \Delta}_j-{\bf u}_j),
\eqno (10)$$
where a is a parameter of the order of the size of the defect and 
$R_{\ell,j}$ 
is the distance between the $\ell^{th}$ and $j^{th}$ defects. This 
equilibrium condition is discussed in more detail in the appendix.
To lowest order in the inter-defect interaction, the approximate solution
for ${\bf u}_{\ell}$ is
$${\bf u}_{\ell}={\bf u}_{\ell}^{0}+[1+(E'a)^{-1}\alpha_{\ell}]^{-1}
\sum_j (E'R_{\ell,j})^{-1}
\alpha_j ({\bf \Delta}_j-{\bf u}^0_j),
\eqno (11a)$$
where 
$${\bf u}_{\ell}^0={E' a\over E' a+\alpha_{\ell}}\Delta_{\ell} \eqno (11b)$$ 
is the zeroth order approximation (i.e., the solution
to Eq. (10) neglecting the second term on the right hand side of the equation).
Since the defects are randomly distributed over the interface,
we can estimate the second term (i.e., the summation over j) on the right hand
side of Eq. (11a) by its root mean square (RMS) average which is estimated by
integrating the square of the summand over the position of the $j^{th}$
defect which is in contact with the substrate and
multiplying by the density of asperities in contact with the substrate
$\rho$. Since the angular integrals only give a factor of order unity,
we need only consider the integral over the magnitude of $R_{\ell,j}$, giving
an RMS value of the sum over $R^{-1}$ of order $[\rho ln(W/a)]^{1/2}$
where here W is
the width of the interface and a is the mean defect size. For $W\approx 1cm$
and $a\approx 10^{-8}cm$, $[ln(W/a)]^{1/2}$
is of order unity. For a defect potential of strength $V_1$, $\alpha_{\ell}
\approx V_1/b^2$, where b is of the order of a lattice constant. If $V_1\approx 
1eV$ and $b\approx 3\times 10^{-8}cm,$ $\alpha\approx 2\times 10^3 dyn/cm^2$. For 
$E'\approx 10^{12}dyn/cm$, $u_{\ell}\approx (\alpha_{\ell}/E'b)\Delta_{\ell}
\approx 0.06\Delta_{\ell}.$ This implies that the elasticity of the solid 
prevents the solid from distorting to any significant degree to accommodate 
the defects at the interface, which implies that we are in the weak pinning 
limit. From the scaling arguments of this section, we conclude that there 
will be no static friction in the macroscopic interface limit. 
For stronger defect potentials and/or smaller values of E', however, it is 
clear that we could also be in the strong pinning limit, in which the 
Larkin length is comparable to a lattice spacing, and hence there is static friction. 
For almost any surface, contact only takes placed at random asperities
of mean size and spacing of the order of microns. In the next section it will 
be argued when one treats the interface on the multi-micron scale as a 
collection of contacting 
asperities, one finds in contrast that it is almost certainly in the strong 
pinning limit.
\endpage
\noindent {\bf III. Static Friction due to Disorder on the Micron Length Scale}

The arguments in the last section seem to imply that weakly interacting disordered
surfaces cannot exhibit static friction. We shall see, however, that
unlike weak atomic scale defects, for which the elastic interaction
between them can dominate over their interaction with the second surface,
for contacting asperities that occur when the problem is studied on the
multi-micron scale, the interaction of two contacting asperities from 
the two different surfaces dominates over the
elastic interaction between two asperities in the same surface. 
It is suggested here that this could be
responsible for the virtual universal occurrence of static friction.
Roughness due to asperities is well described by
 the Greenwood-Williamson (GW) model[15], in which there
are assumed to be elastic spherically shaped asperities
on a surface with an exponential or Gaussian height distribution
in contact with a rigid flat substrate,
especially for relatively light loads. 
Volmer and Nattermann presented a possible approximate way of accounting 
for dry friction[17]. Their discussion of static
friction, however, is not qualitatively different from that of Ref. 16. 
In the GW model, the 
total contact area is given by
$$A_c=2\pi\sigma NR_c\int_h^{\infty} ds \phi (s) (s-h), \eqno (12)$$
where $\phi (s)$ is the distribution of asperity heights s, in units of a length
 scale $\sigma$ associated with the height distribution, $R_c$ is the 
 radius of curvature of a typical asperity and h is the distance 
of the bulk part of the sliding solid from the flat surface in which it is in 
contact, measured in units of $\sigma$. Since the force of static friction 
exerted on a single asperity is expected to be equal to the product of the 
contact area and a shear strength for the 
interface, it is proportional to this quantity. 

The number of contacting asperities per unit surface
area is given by
$$\rho (h)=(N/A)\int_h^{\infty} ds \phi (s). \eqno (13),$$
where A is the total surface area and N is the total number of asperities
whether in contact with the substrate or not. The integral in
Eq. (12) divided by the integral in Eq. (13), which is proportional to the
contact area per asperity and the square root of the integral in
Eq. (13) are plotted as a function of the load, which is given in this
model by
$$F_L=(4/3)E'N(R_c/2)^{1/2}\sigma^{3/2}\int_h^{\infty} ds \phi (s) (s-h)^{3/2},
\eqno (14)$$
in Fig. 1. 
A Gaussian distribution is assumed here for $\phi (s)$ (i.e.,
$\phi (s)=(2\pi)^{-1/2}e^{-s^2/2}$).
Since the square root of Eq. (13) drops to zero
in the limit of vanishing load, whereas Eq. (12) divided by Eq. (13)
approaches a nonzero
value, this implies that the interface will approach the strong pinning
regime (i.e., the regime in which the asperity-substrate interaction
dominates over the inter-asperity interaction) in the limit of vanishing load.

Let us now apply the equilibrium conditions expressed in 
Eqs. (10) and (11), used in the last section, to the asperities.
In this section, a is taken to be a parameter of the order of the 
size of the asperity.
Since the contacting asperities are randomly distributed over the interface,
we can again estimate the second term (i.e., the summation over j) on the right hand
side of Eq. (10) by its root mean square (RMS) average which is estimated by
integrating the square of the summand over the position of the $j^{th}$
asperity, which is in contact with the substrate, over its position and
multiplying by the density of asperities in contact with the substrate
$\rho$,  giving
an RMS value of the sum over $R^{-1}$ of order $[\rho ln(W/a)]^{1/2}$
where here W is
the length of the interface and a is the asperity size. For $W\approx 1cm$
and $a\approx 10^{-4}cm$, $[ln(W/a)]^{1/2}$
is of order unity. The energy of the system can
be written as
$$(1/2)\sum_j \alpha_j |{\bf \Delta}_j-u_j|^2+$$
$$(1/2)E'\sum_j\int d^3 r [|\nabla {\bf G}({\bf r})|
(\alpha_j |{\bf\Delta}_j-{\bf u}_j)|]^2 \eqno (15)$$
It follows from Eqs. (10,11,15) that the two lowest order nonvanishing
terms in an
expansion of the energy of the system in powers of $\rho^{1/2}$ are the
zeroth and first order ones. The shearing of the junction at the area of 
contact of two asperities involves the motion of two atomic planes relative to 
each other, and hence the distance over
which the contact potential varies must be of the order of atomic distances.
Then, if we denote the width of the asperity contact potential well 
(i.e., the length scale over which the contact potential varies) by b,
of the order of atomic spacings, we must choose a typical value
for $\alpha$ such that $\alpha b$ is of the order of the shear rupture
strength of the asperity contact junction ($\approx E'\pi a^2$). 
Thus, $\alpha>>E'a$, since $a>>b$. Then, applying Eq. (11b) to the contacting 
asperities, we find that ${\bf u_0}\approx {\bf \Delta_{\ell}}$, i.e., the 
contacting asperities lie at the minima of the contact potential. This is very 
easy to understand. Since the contact potential varies over distances of the 
order of an atomic spacing, the asperities can all sink very close to their 
contact potential minima by moving a distance of the order of an atomic 
spacing, with negligible cost in elastic potential energy. 
Zeroth order in the asperity density in Eq. (15) is of the order of
$\alpha\Delta^2$, where $\alpha$ is a typical value of $\alpha_j$,
and $\Delta$ is a typical value of $\Delta_j$.
The term linear in $\rho^{1/2}$ is easily shown to be of the order of
$E'a^2\Delta^2 \rho^{1/2}$, independent of $\alpha$ to zeroth 
order in $E'a/\alpha$. Since it depends on $\rho$ it
represents an interaction energy between the asperities.

Let us now give sample numerical values for
some of the quantities which occur in the
application of the GW model to this problem. Following Ref. 16, we 
choose $\sigma=
2.4\times 10^{-4}mm$ and $R_c=6.6\times 10^{-2} mm$, and assume that there
is a density $\rho$ of $4.0\times 10^3$ asperities/$mm^2$.
Then by performing the integrals
in Eqs. (12-14), we find that for $F_L/A=3.98\times 10^{-4} N/mm^2$,
where A is the apparent area of the interface, the total contact area divided
by A is $3.03\times 10^{-5}$, and the contact area per asperity from
the ratio of Eqs. (12) and (13) is $2.44\times 10^{-5} mm^2$.
Also, $\rho (h)^{1/2}$, which is equal to
the square root of Eq. (13) is $1.11 mm^{-1}$. The mean interasperity
interaction
force is approximately equal to the derivative of the first order term in
$\rho^{1/2}$ given below Eq. (15) with respect to $\Delta$
or $E'a^2\rho (h)^{1/2}\Delta$, where a is taken as the square
root of the mean contact area per asperity divided by $\pi$.
The mean strength of the force acting on an asperity,
due to the solid with which it is in contact, will be estimated by the product
of its contact area and the shear rupture strength $E_r$.
Then, the condition for the latter
quantity to dominate over the asperity-asperity interaction,
$E_r\pi a^2>E' 4\pi a^2\rho^{1/2}\Delta$ or
$E_r/E'>4\rho (h)^{1/2}\Delta$, is easily satisfied by the above
calculated quantities since the right hand side is $4\times 10^{-7}$ 
(since $\Delta$ is of the order of the potential well width or 
$10^{-8}cm$ ) and
the left hand side is of order unity because for an asperity, 
which is typically dislocation free, $E_r$ is comparable to the 
shear modulus, which is of the order of E'.

For higher loads since the density of contacting asperities increases, 
 the system appears to move towards the "weak pinning" limit, the latter 
conclusion is most likely incorrect, however, because it does not take
into account the fact that the distribution of asperity heights contains
asperities which are much higher than average. These asperities will be
compressed much more than a typical asperity, making the friction force on
them considerably larger than average. Since the probability of such
asperities occurring is relatively small, however, they will be typically
far apart, putting them in the strong pinning limit (i.e., each one lies 
in the bottom of its potential well). For example, the
probability of the ratio of an asperity height to $\sigma$
being greater than h by a value $h_L$ is
$$P(h_L)=\int_{h+h_L}^{\infty} ds \phi (s), \eqno (16)$$
whose mean height and hence contact area is proportional to
$$P(h_L)^{-1}\int_{h+h_L}^{\infty} ds \phi (s) (s-h). \eqno (17)$$
These two quantities are plotted in Fig. 2.
It is seen that even for $h_L$ only equal to 1/2 [corresponding
to an asperity height
equal to $(1.5\sigma)$], Eq. (17)
remains larger than the square root of Eq. (16). It is most likely only 
possible to apply the GW model to quite light loads anyway. For higher loads, 
plasticity becomes important[16]. 

Although it has been argued here that the GW model
predicts the occurrence of a sufficiently dilute concentration of
asperities with strong enough forces acting on them due to the second
solid to consider the asperities to be essentially uncorrelated (i.e., their 
Larkin lengths are comparable to their separation), this still
does not necessarily guarantee that there will be static friction, 
since it has been argued that even for uncorrelated asperities, static
friction will only occur if the asperities exhibit multistability[7,9,17]. The
condition for multistability to occur at an interface[9], namely that the
force constant due to the asperity contact potential be larger than
that due to the elasticity of the asperity ($\approx E'a$), however, is
satisfied, as discussed under Eq. (15).

In conclusion, when one considers atomically smooth surfaces,
arguments based on Larkin domains indicate that the disorder at an interface
between two weakly interacting nonmetallic elastic solids in contact 
will not result in static
friction. When one applies such arguments to the distribution of asperities
that occur on multimicron length scales, however, one finds that the asperities
are virtually always in the "strong pinning regime," in which the the Larkin
domains are comparable in size to a single asperity. This accounts for the
fact that there is almost always static friction. Muser and Robbins' idea
[2],however, is not invalidated by this argument. Their result will still apply 
for a smooth crystalline interface. It will also apply in the present context
to the contact region between two asperities, implying that for a clean
interface the shear force between contacting asperities is proportional to the
square root of the contact area. The GW model predicts for this case
that the average force of friction is proportional to the 0.8 power of the
load[16], but this load dependence is not significantly different from 
when the asperity contact force is proportional to the contact area. 
This is illustrated in Fig. 3, where the both the integral over s 
in Eq. (12) and the integral,
$$\int_h^{\infty} ds \phi (s) (s-h)^{1/2} \eqno (18)$$
are plotted as a function of the load. This quantity, and hence the static 
friction, are approximately proportional to the 0.8 power of the load. 
Furthermore, some simple arguments show that the although the Muser-Robbins
[2] picture, when the effects of asperities cosidered in the present work are 
taken into account, does not allow one to conclude that there will be no 
static friction for clean surfaces, it does predict that the static friction 
for clean surfaces is much smaller than what is normally observed. The 
argument is as follows: If the interface between two asperities is either 
in the strong pinning limit or using the Muser-Robbins[2] picture, it contains 
a sub-monolayer of mobile molecules, the force of static friction per 
asperity is given by 
$$F_s/N=E_r <A_c>, \eqno (19)$$
where one expects for the shear rupture strength at the asperity contact 
region, $E_r$,
$$E_r\approx cV_0/b^3, \eqno (20)$$
where c is the concentration of defects at the interface, $V_0$ is the 
strength of the defect potential and b is of the order of a lattice 
constant. Using the sample parameters given earlier in this section, 
we obtain as an estimate of the static friction coefficient
$$\mu_s=F_s/F_L\approx 0.1. \eqno (21)$$
According to the Muser-Robbins argument, for clean surfaces, 
Eq. (19) is replaced by
$$F_s/N\approx E_r <A_c (b^2/c A_c)^{1/2}>=E_r b<A_c^{1/2}>/c^{1/2} \eqno (22)$$
which when one again substitutes the sample parameters given earlier 
in this section gives
$$\mu_s\approx 10^{-5}. \eqno (23)$$
On the basis of this argument, one concludes that the ideal static friction 
coefficient between clean, weakly interacting surfaces in the light load limit, 
is much smaller than what one typically observes.
\endpage
\noindent {\bf Acknowledgements}

I wish to thank the Department of Energy (Grant DE-FG02-96ER45585) for their 
financial support. I also wish to thank R. Markiewicz for useful discussions.
\endpage
\noindent {\bf Appendix A-Equilibrium Condition for Pinning Centers at an 
Interface}

The interaction forces between two solids in contact act at various points 
along the interface. For atomically flat surfaces, they are expected to act 
at the points of contact of the atoms of both solids at the interface. Since 
it has been established, however, that weakly interacting perfect 
incommensurate surfaces exhibit no static friction [1-3], we expect that any 
static friction that occurs is due to defects. Therefore, let us consider the 
interaction of point defects, randomly distributed over the interface. We must 
consider both the potential of interaction of a defect of one surface with 
an atom on the second surface and the elastic energy cost that one must 
pay when the solid distorts in order to minimize the interfacial and elastic 
potential energies. If ${\bf f_j}$ represents the force acting on an atom 
at site j in the solid due to its interaction with a second solid with which 
it is in contact, the displacement of the atom at the $\ell^{th}$ lattice site 
${\bf u_{\ell}}$ is given by 
$${\bf u_{\ell}}={\bf G_{\ell,j}}\cdot {\bf f_j}, \eqno (A1)$$
where 
$${\bf G_{\ell,j}}={\bf D_{\ell,j}}^{-1} \eqno (A2)$$
where ${\bf D_{\ell,j}}$ is the dynamical matrix[14,15]. For implicity, we 
assumed in Eq. (10) that ${\bf f_j}$ has the form
$${\bf f_j}=\alpha_j ({\bf\Delta_j}-{\bf u_j}). \eqno (A3)$$
Then Eq. (A1) can be written as 
$${\bf u_{\ell}}={\bf G_{\ell,\ell}}\cdot 
\alpha_{\ell} ({\bf\Delta_{\ell}}-{\bf u_{\ell}})+
\sum_{j\ne\ell} {\bf G_{\ell,j}}\cdot 
\alpha_j ({\bf\Delta_j}-{\bf u_j}). \eqno (A4)$$
Following the discussion in Ref. 14, we find that 
$${\bf G_{\ell,j}}=v(2\pi)^{-3}\sum_{\gamma}\int d^3 k 
e^{i{\bf k}\cdot ({\bf R_{\ell}}-{\bf R_j})}
{\hat\epsilon_{\bf k}^{\gamma}\hat\epsilon_{\bf k}^{\gamma}\over 
m\omega^2_{\gamma} ({\bf k})}, \eqno (A5)$$
where m is the mass of an atom in the solid, $\hat\epsilon_{\bf k}^{\gamma}$ 
is the unit vector 
which gives the polarization of the $\gamma^{th}$ phonon 
mode of wavevector ${\bf k}$, ${\bf R_j}$ is the 
location of the $j^{th}$ atom and v is the unit cell volume. 
In order to simplify the problem, let us replace the tensor 
$\hat\epsilon_{\bf k}^{\gamma}\hat\epsilon_{\bf k}^{\gamma}$, 
by the unit tensor, which should give results of the correct order of 
magnitude . Then Eq. (A5) becomes when the integral 
over k is done in the Debye approximation
$${\bf G_{\ell,\ell}}\approx {9\over mc^2k_D^3}
\int_0^{k_D} dk{sin(kR)\over k}, \eqno (A6)$$
where $R=|{\bf R_{\ell}}-{\bf R_j}|$, c is the mean sound 
velocity and we have used the fact that 
the Debye wavevector $k_D$ is related to v by
$$v(2\pi)^{-3}4\pi k_D^3/3=1. \eqno (A7)$$
In Ref. 14, it is shown that the elastic constants are given 
by
$$-(1/2v)\sum_{\bf R} {\bf R}{\bf D}({\bf R}) {\bf R}. \eqno (A8)$$
The magnitude of a typical value of an elastic constant E is given by
$$E=(1/2)(2\pi)^{-3}b^2\sum_{\gamma}\int d^3 k m\omega_{\gamma}^2 ({\bf k}). 
\eqno (A9)$$
When Eq. (A9) is evaluated in the Debye approximation, we obtain 
$$E=(3/10\pi^2)mc^2 k_D^5 b^2. \eqno (A10)$$
For $k_D R>>1$, we find using Eqs. (A6) and (A10), taking 
$k_D b\approx\pi$ 
$${\bf G_{\ell,j}}=(E' R)^{-1},\eqno (A11)$$
where $E'=(40/9)E$. For $k_D R<<1$, 
$$G\approx (E' b)^{-1} \eqno (A12).$$
Then, from Eqs. (A1), (A3), (A11) and (A12), we obtain Eq. (10).

The equilibrium condition expressed in Eq. (10) can also be applied to 
an interface for which the contact takes place only at a dilute 
concentration of randomly placed contacting asperities, giving 
for the displacement at a point on the $\ell^{th}$ asperity 
$${\bf u_{\ell}}=\sum_j \int d^2 r'_j 
(E'|{\bf r_{\ell}}-{\bf r_j}-{\bf r'_j}|)^{-1}{\bf p}({\bf r'_j}),
\eqno (A13)$$
where ${\bf r_j}$ is the location of a central point in the contact 
area of the $j^{th}$ asperity, ${\bf r'_j}$ gives the location 
of an arbitrary point on this asperity relative to ${\bf r_j}$ 
and ${\bf p}({\bf r'_j})$ is the shear stress at the point 
${\bf r'_j}$. We have replaced the summation over atomic positions 
in Eq. (A1) by the integral over ${\bf r'_j}$ over the contact area 
of the $j^{th}$ asperity. In the dilute asperity  
limit, in which $|{\bf r_{\ell}}-{\bf r_j}|>>r'_j$, Eq. (A13) 
is to a good approximation
 $${\bf u_{\ell}}=\sum_{j\ne\ell}  
(E'|{\bf r_{\ell}}-{\bf r_j}|)^{-1}{\bf f_j}+
\int d^2 r'_{\ell}(E'|{\bf r_{\ell}}+{\bf r'_{\ell}}|)^{-1}
{\bf p}({\bf r'_{\ell}}),
\eqno (A14)$$
where ${\bf f_j}=\int d^2 r'_j {\bf p}({\bf r'_j})$, where the 
range of integration is over the contact area of the $j^{th}$ 
asperity. For simplicity, we may replace ${\bf p}({\bf r'_{\ell}})$ 
in the integral over $r'_{\ell}$ by its average value, denoted by 
$\pi a^2 {\bf f_{\ell}}$. Then we need to 
estimate the integral
$$\int d^2 r'|{\bf r}+{\bf r'}|^{-1} \eqno (A15)$$
where ${\bf r}$ denotes a point on the $\ell^{th}$ asperity 
and the integral runs over the contact area of this asperity. 
Taking the contact area to be a circle of radius a, this 
integral can easily be shown to be equal to 
$$4\int_0^{\infty} dr' r'(r+r')^{-1} K({4rr'\over r+r'}), \eqno (A16)$$
where K(k) is the complete elliptic function. It has a 
logarithmic singularity at $r'=r$, which is integrable, and is of 
order 1 away from the singularity. Consequently, the integral is 
of order a and we obtain a contribution of order 
$(E' a)^{-1}{\bf f_{\ell}}$ for the last term in Eq. ((A14). 
If for simplicity, we replace ${\bf f_j}$ by 
$\alpha_j ({\bf \Delta_j}-{\bf u_j})$, as was done in section 
III, and we obtain the equilibrium condition for the asperities 
used in that section. 
\endpage
\noindent {\bf References} 
\bigskip\noindent
[1]  J. E. Sacco and 
J. B. Sokoloff, Phys. Rev. B 18 (1978) 6549. \nextline
[2] G. He, M. H. Muser and M. O. Robbins, Science 284 (1999) 1650 ; 
M. H. Muser and M. O. Robbins, Phys. Rev. B 61 (2000) 2335; 
M. H. Muser, L. Wenning and M. O. Robbins, Phys. Rev. Lett. 86, 
1295 (2001). \nextline
[3] S. Aubry, in Solitons and Condensed Matter, ed. A. R. Bishop and
T. Schneider (Springer, New York, 1978), p. 264. \nextline
[4] J. B. Sokoloff, 
Phys. Rev. B51 (1995) 15573; J. B. Sokoloff and M. S. Tomassone, 
 Phys. Rev. B 57 (1998) 4888. \nextline
[5]  M. S. Tomassone and J. B. Sokoloff, Physical Review B60 (1999) 4005.
\nextline
[6] G. A. Tomlinson, Phil. Mag. 7 (1929) 905. \nextline
[7] C. Caroli and Ph. Nozieres, European Physical Journal B4 (1998) 233; 
Physics of Sliding Friction, edited by B. N. J. Persson and E. Tosatti, 
NATO ASI Series E: Applied Sciences, Vol. 311 (Kluwer Acad. Publ., 
Dordrecht, 1996). \nextline
[8] D. S. Fisher, Phys. Rev. B31 (1985) 1396; Phys. Rev. Lett. 50 (1983)
1486.\nextline
[9] B. N. J. Persson and E. Tosatti in Physics of Sliding Friction, ed.
B. N. J. Persson and E. Tosatti (Kluwer Academic Publishers, Boston, 1995),
p. 179;  V. L. Popov, Phys. Rev. Lett. 83 (1999) 1632.\nextline
[10]  M. L. Falk and J. S. Langer, Phys. Rev. E 57 (1998) 7192. \nextline
[11] A. I. Larkin and Yu. N. Ovchinikov, J. Low Temp. Phys. 34 (1979) 409.
\nextline
[12] H. Fukuyama and P. A. Lee, Phys. Rev. B17 (1977) 535; 
P. A. Lee and T, M. Rice, Phys. Rev. B19 (1979) 3970.\nextline
[13] J. B. Sokoloff, Phys. Rev. B23 (1981) 1992. \nextline
[14] N. W. Ashcroft and N. D. Mermin, "Solid State Physics" (Saunders College, 
Philadelphia, 1976), pp. 443-444. \nextline
[15]  L. D. Landau and E, M. Lifshitz, Theory of Elasticity, (Pergamon
Press, New York, 1970), p. 30. \nextline
[16]  J. A. Greenwood and J. B. P. Williamson, Proc. Roy. Soc. A295 (1966) 3000; 
J. I. McCool, Wear 107 (1986) 37 . \nextline
[17] A. Volmer and T. Nattermann, Z. Phys. B104 (1997) 363. 
\endpage
\noindent {\bf Figure Captions} 
\medskip
\noindent 1. The curve which is lower at the right is a plot of the integral in 
Eq. (12) divided 
by the integral in Eq. (13) and the curve which is higher on 
the right is a plot of the square root of the integral in Eq. (13) versus 
the integral in Eq. (14). All quantities are dimensionless.\nextline
2. Eq. (17) (the higher curve) and the square root of Eq. (16) 
(the lower curve) are plotted versus the load Eq. (14) 
divided by $(4/3)E(b/2)^{1/2}\sigma^{3/2}$. All quantities are dimensionless. 
\nextline
3. The integral over s in Eq. (12) (the lower curve) and the integral over s 
in Eq. (18) (the upper curve) are plotted as a function of the integral over 
s in Eq. (14), which is proportional to the load. All quantities are 
dimensionless.
 \end

\noindent {\bf Appendix A-Pinning for Dilute Asperities}

When asperity-asperity interaction of each of the sliding solids dominates 
over asperity interaction across the interface, we saw that there is no static 
friction for two three dimensional solids in contact in the thermodynamic 
limit. In the dilute asperity limit, in which the asperity-asperity interaction 
can be taken as vanishingly small, in contrast, there should be static 
friction. In the latter limit, the asperities act independently. When the 
solids are brought into contact the lattice distorts so that the interaction 
energy of each asperity with the asperity of the second solid with which it 
is in contact plus the elastic energy due to the distortion reaches a minimum. 
The amount of distortion which takes place at each contacting asperity depends 
on how far from equilibrium the contacting asperities are before the distortion 
takes place. When an external stress is applied to try to slide the solids 
relative to each other, the resulting force which acts on each asperity tries 
to push it out of its potential minimum. Since each asperity is at a potential 
minimum, however, there will clearly be an opposing force which must be 
overcome in order to get the contacting asperities to slide relative to each 
other. Thus there will always be static friction in the dilute asperity limit. 

To illustrate this, consider a distribution of asperities at an interface, 
each one located at a random position $\vec\Delta$ with respect to the minimum of 
a spherically symmetric potential V(r). If the solid does not distort at all, 
the mean force on an asperity is given by 
$$-\Delta_0^{-2}\int d^2 r \nabla V(r)=0,$$
where $\Delta_0$ is the range of one component of $\vec\Delta$. When the 
distortion $\vec u$ takes place, the total potential energy of a single 
asperity in the presence of an applied force $\vec f_0$ is given by 
$$(1/2)\alpha u^2+V(|\vec\Delta+\vec u|)-\vec f_0\cdot\vec u.$$
The condition for a minimum of this energy is
$$\alpha\vec u+(\vec u/\Delta)V'-
(\vec\Delta/\Delta)\vec u\cdot\vec (\vec\Delta/\Delta)V''=
\vec f_0-(\vec\Delta/\Delta)V',$$
where $V'=dV(r)/dr|_{r=\Delta}$ and $V''=d^2V(r)/dr^2|_{r=\Delta}$. 
If we choose the coordinate axes such that the x-axis is along the direction 
of $\vec f_0$, the pair of simultaneous equations for the x and y components 
of $\vec u_0$, represented by the above equation, can easily be solved, 
giving
$$u_x=[(f_0-V'\Delta_y/\Delta)(\alpha+V'/\Delta-V'' (\Delta_y/\Delta)^2)-
(V''\Delta_x\Delta_y^2/\Delta^3]/D,$$
$$u_y=[(f_0-V'\Delta_x/\Delta)(V''\Delta_x\Delta_y/\Delta^2)-
(\alpha+V'/\Delta-V''\Delta_x^2/\Delta^2)V'\Delta_y/\Delta]/D,$$
where
$$D=(\alpha+V'/\Delta-V''\Delta_x^2/\Delta^2)
(\alpha+V'/\Delta-V''\Delta_y^2/\Delta^2)-(V''\Delta_x\Delta_y/\Delta^2)^2.$$
Since D generally does not vanish, there always exists a solution for $u_x$ 
and $u_y$ for sufficiently small values of $f_0$. The averages over the 
above expressions for $u_x$ and $u_y$ over $\delta_x$ and $\delta_y$ 
give 
$$u_x=f_0 <D^{-1}>,$$
and
$$u_y=0.$$
Thus, if the elastic interaction between the asperities in a solid is 
neglected, there will always be static friction. 

[1]. J. Krim, Comments Cond. Mat. Phys. 17, 263 (1995). \nextline
[2]. C. M. Mate, G. M. McClellan, R. Erlandsson and S. Chang, Phys. Rev. Lett. 
59, 1942 (1987); G. Neubauer, S. R. Cohen, G. M. McClelland, D. Horne and 
C. M. Mate, Rev. Sci. Instrum. 61, 2296 (1990); C. M. Mate, Phys. Rev. Lett. 
68, 3323 (1992); C. M. Mate, Wear 168, 17 (1993). \nextline
[3]. G. Reiter, A. L. Demirel, J. Peanasky, L.L. Cai and S. Granick, J. Chem. 
Phys. 101, 2606 (1994); S. Granick, A. L. Demirel, L. Cai, and J. Peanansky, 
Isral Journal of Chemistry 35, 75 (1995); A. L. Demirel and S. Granick in 
"Physics of Sliding Friction," ed. B.N.J. Persson and E. Tosatti (Kluwer 
Academic Publishers, Boston, 1996). \nextline
[4]. J. Krim, D. H. Solina and R. Chiarello, Phys. REv. Lett. 66, 181 (1991); 
J. Krim and c. Daly in "Physics of Sliding Friction," ed. B.N.J. Persson and 
E. Tosatti Kluwer Academic Publishers, Boston, 1996). \nextline
[5]. J. B. Sokoloff, Wear 167, 59 (1993); Phys. Rev. B42, 760, 6745 (erratum) 
(1990); J. Appl. Phys. 72, 1262 (1992); Phys. REv. 47, 6106 (1992); Journal 
of Physics-Condensed Matter 10, 9991 (1998). \nextline
in Physics of Sliding Friction, ed. B. N. J. Person and E. Tosatti,
NATO ASI Series, Series E: Applied Sciences-Vo. 311 (Kluwer Academic
Publishers, Dordrecht, 1996), p. 217. \nextline
[6]. M. Cieplak, E. D. Smith and M.O. Robbins, Science 265, 1209 (1994); E. D. 
Smith, M. O. Robbins and M. Cieplak, Phys. Rev. B 54, 8252 (1996); P. A. 
Thompson, G. S. Grest and M. O. Robbins, Phys. Rev. Lett. 68, 3448 (1992); 
P. A. Thompson, M. O. Robbins and G. S. Grest, Israel Journal of Chemistry 35, 
93 (1995); M. S. Tomassone, J. B. Sokoloff, A. Widom and J. Krim, Phys. Rev. 
Lett. 79, 4798 (1997); B.N.J. Persson and A. Nitzan, Surf. Sci. 367, 261 (1996);
 B. N. J. Persson and A. Nitzan, Surf. Sci. 367, 261 (1996); A. Liebsch, 
S. Goncalves and M. Kiwi, to be published; J. A. Harrison, C. T. White, R. J. 
Colton and W. D. Brenner, Thin Solid Films 260, 205 (1995); J. A. Harrison, R. 
J. Colton, C. T. White and D. W. Brenner, Phys. Rev. B46, 9700 (1992); M. D. 
Perry and J. A. Harrison, Langmuir 12, 4552 (1996); J. Phys. Chem. B 101, 1364 
(1997); J. A. Harrison, S. J. Stuart and M. D. Perry, Proc. Workshop on 
Tribology Issues and Opportunities in MEMS (Kluwer Academic Publishers, 1998); 
J. A. Harrison, S. J. Stuart and D. W. Brenner, in Handbook of 
Micro/Nanotribology, ed. B. Bhushan (CRC Press, Boca Raton, FL 1998); J. A. 
Harrison and S. S. Perry, MRS Bulletin 23, 27 (1998). \nextline
[7]. M. G. Rozman, M. Urbakh and J. Klafter, Phys. Rev. Lett. 77, 683 (1996); 
Phys. Rev. E 54, 6485 (1996); V. Zaloj, M. Urbakh and J. Klafter, Phys. Rev. 
Lett. 81, 1227 (1998); Y. Braiman, F. Family and H. G. E. Hentschel, Phys. Rev. 
B55, 5491 (1997); Phys. Rev. 53, 3005 (1996); O. M. Braun, T. Dauxois, M. V. 
Paliy and M. Peyrard, Phys. Rev. Lett. 78, 1295 (1997); Phys. Rev. E 55, 3598 
(1997); M.V. Paliy, O. M. Braun, T. Dauxois, Bambi Hu, Phys. Rev. E56, 4025 
(1997). \nextline
[8]. J. B. Sokoloff, Phys. Rev. Lett. 71, 3450 (1993); Phys. Rev. B 52, 7205 
(1995). \nextline 
[9]. J. B. Sokoloff, Phys. Rev. B51, 15573 (1995); B. N. J. Persson and A. I. 
Volokitin, J. Phys.:Condens. Matter 9, 2869 (1997). \nextline
[10]. D. S. Fisher, Phys. Rev. B31, 1396 (1984). \nextline 
[11]. A. Dayo, W. Alnasrallah and J. Krim, Phys. Rev. Lett. 80, 1690 (1998). 
\nextline
[12] B.N.J. Persson, Phys. Rev. B 44, 3277 (1991);
B.N.J. Persson and A. Volokitin, J. Chem. Phys. 103, 8679 (1995); 
J. B. Sokoloff, Phys. Rev. B 52, 5318 (1995); L. S. Levitov, Europhysics 
Letters 8, 499 (1989). \nextline
[13] M. S. Tomassone and A. Widom, Phys. Rev. B 56, 4938 (1997); Am. J. Phys. 
65, 1181 (1997); T. H. Boyer, Phys. Rev. A 9,68 (1974). \nextline
[14] C. Wang and R. Gomer, Surf. Sci. 91, 533 (1980); surf. Sci. 84, 329 (1979); 
Surf. Sci. 74, 389 (1979); P. W. Palmberg, Surf. Sci. 25, 598 (1971).; T. 
Engel and R. Gomer, J. Chem. Phys. 52, 5572 (1979); J. C. P. Mignolet, J. Chem. 
Phys. 21, 1298 (1953); Discussions Faraday Soc. 8, 105 (1950). \nextline
[15]. N. W. Ashcroft and N. D. Mermin, "Solid State Physics" (Saunders College, 
Philadelphia, 1976), pp. 340-342. \nextline
[16]. J. Krim and A. Widom, Phys. Rev. B38, 12184 (1988). \nextline
{17} M.O. Robbins and E. D. Smith, Langmuir 12, 4543 (1996). \nextline
{18} H. Fukuyama and P. A. Lee, Phys. Rev. B 17, 535 (1978). \nextline
[19]. G. A. Tomlinson, Phil. Mag. 7, 905 (1929). \nextline
[20]. C. Caroli and P. Nozieres, in "Physics of Sliding Friction," ed. 
B. N. J. Persson and E. Tosatti (Kluwer Academic Publishers (Boston, 1996), 
p. 27;  C. Caroli and P. Noziere, Eur. Phys. Lett. 44, 233 (1998); 
T. Baumberger, P. Berthoud and C. Caroli (preprint). \nextline
[21] M. L. Falk and J. S. Langer, Phys. Rev. E 57, 7192 (1998). \nextline
[22] J. E. Sacco, J. B. Sokoloff and A. Widom, Phys. Rev. B20, 5071 (1979).
\nextline
[23] X. Xiao, J. Hu, D. H. Cjaarych and M. Salmeron, Langmuir 12, 235 (1996); 
Q. Du, X.-d. Xiao, D. H. Charych, F. Wolf, P. Frantz, Y. R.Shen and M. 
Salmeron, Phys. Rev. 51, 7456 (1995); M. Salmeron, Materials Research 
Society Bulletin (May, 1993), 593; G.-yu Liu and M. B. Salmeron, Langmuir 10, 
367 (1994); Xu-Dong Xiao, Gang-yu Liu, D. H. Charych and M. Salmeron, Langmuir 
11, 1600 (1995); Gang-yu Liu, P. Fenter, C. E. D. Chidsey, D. F. Ogletree, 
P. Eisenberger and M. Salmeron, J. Chem. Phys. 101, 4301 (1994); M. Salmeron, 
G. Neubauer, A. Folch, M. Tomitori, D. F. Ogletree and P. Sautet, 
Langmuir 9, 3600 (1993); A. Lio, D. H. Charych and M. Salmeron, J. Phys. 
Chem. B 101, 3800 (1997).  \nextline
[24] E. Barrena, S. Kopta, D.F. Ogletree, D. H. Charych and M. Salmeron, 
Phys. Rev. Lett. 82, 2880 (1999). \nextline
[25] J. Hautman and M. L. Klein, J. Chem. Phys. 91, 4994 (1989); j. Chem. 
Phys. 93, 7483 (1990); J. P. Bareman, G. Cardini and M. L. Klein, Phys. Rev. 
Lett. 60, 2152 (1988); K. J. Tupper, R. J. Colton and D. Brenner, Langmuir 10, 
2041 (1994). \nextline
[26] J. H. Weiner and W. T. Sanders, Phys. Rev. 134, A1007 (1964); 
V. L. Popov, Phys. Rev. Lett. 83, 1632 (1999).

\begin{figure}
\centerline{
\vbox{ \hbox{\epsfxsize=4.5cm \epsfbox{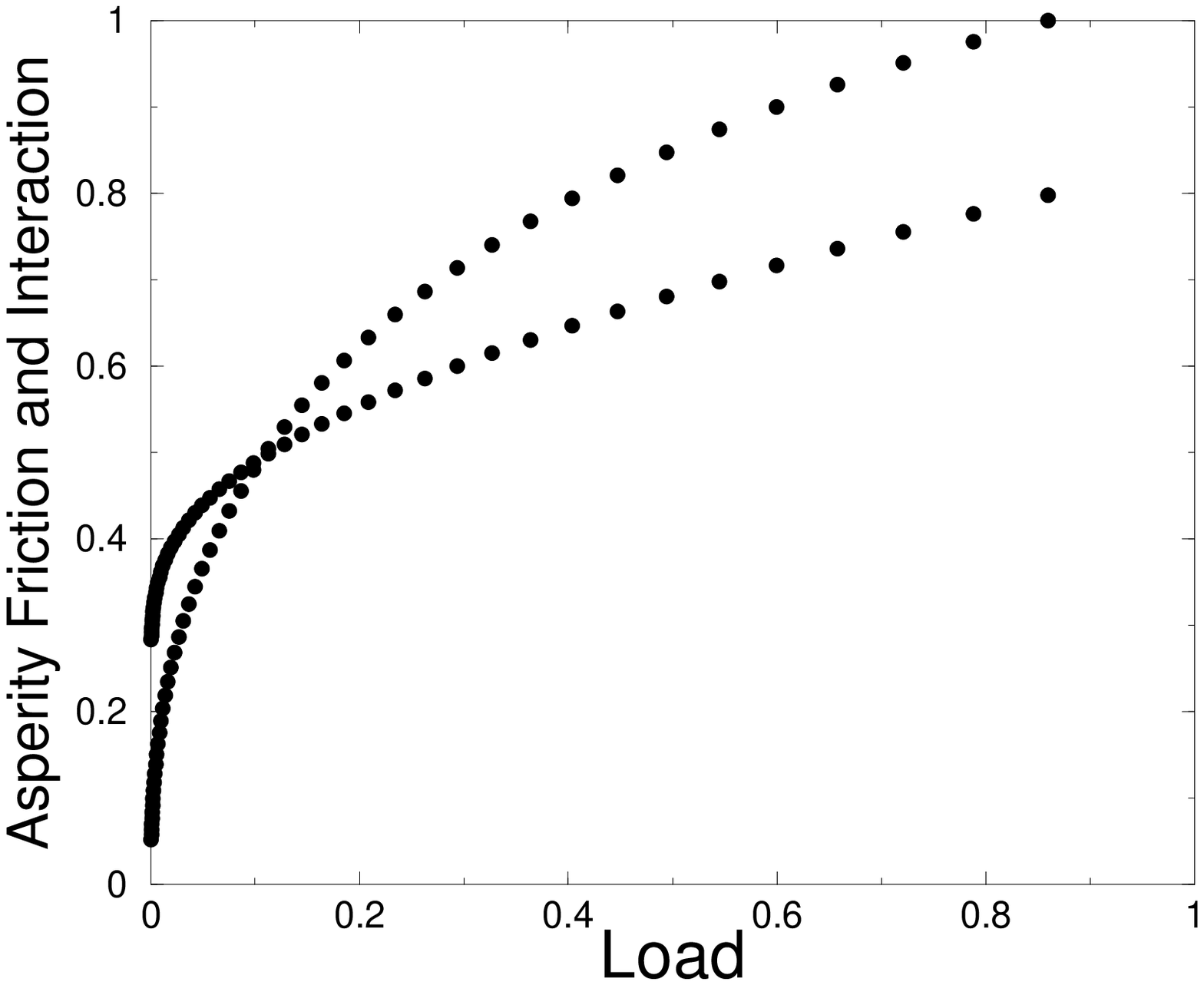}}
       \vspace*{1.0cm}
        }
}
\caption{The curve which is lower at the right is a plot of the integral in
Eq. (4) divided
by the integral in Eq. (8) and the curve which is higher on
the right is a plot of the square root of the integral in Eq. (8) versus
the integral in Eq. (9).
All quantities are dimensionless.}
\label{Fig1}
\end{figure}

\begin{figure}
\centerline{
\vbox{ \hbox{\epsfxsize=4.5cm \epsfbox{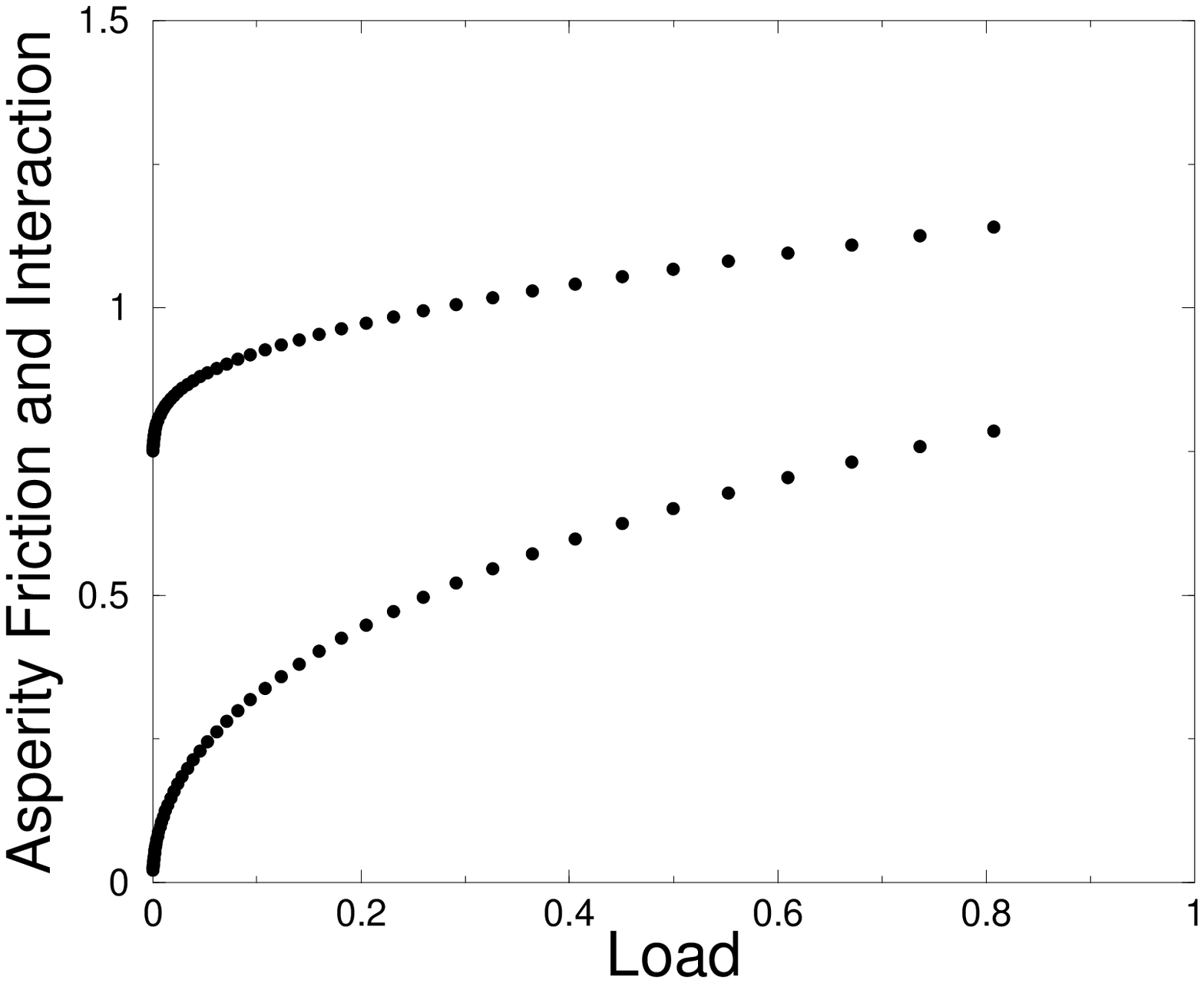}}
       \vspace*{1.0cm}
        }
}
\caption{Eq. (11) (the higher curve) and the square root of Eq. (10)
(the lower curve) are
plotted versus the load Eq. (9)
divided by $(4/3)E(b/2)^{1/2}\sigma^{3/2}$. All quantities are dimensionless.}
\label{Fig2}
\end{figure}